\documentclass{aa}  

\usepackage{graphicx}
\usepackage{tabularx}
\usepackage{txfonts}
\usepackage{xcolor}
\usepackage{soul}
\setstcolor{red}
\def\gtorder{\mathrel{\raise.3ex\hbox{$>$}\mkern-14mu
    \lower0.6ex\hbox{$\sim$}}}
\def\ltorder{\mathrel{\raise.3ex\hbox{$<$}\mkern-14mu
    \lower0.6ex\hbox{$\sim$}}}
\begin{document}

   \title{Feedback from intermediate mass black holes on dwarf galaxy morphology at $z=2$}


   \author{Da Bi
          \inst{1}
          \and
          Dominik R.G. Schleicher\inst{1}\inst{2}
          \and
          Andr\'es Escala\inst{3}
          }

   \institute{Departamento de Astronom\'ia, Facultad Ciencias F\'isicas y Matem\'aticas,    Universidad de Concepci\'on, Av. Esteban Iturra s/n Barrio Universitario, Casilla 160-C, Concepci\'on, Chile\\
              \email{bida@udec.cl}
        \and
        Dipartimento di Fisica, Sapienza Universit´a di Roma, Piazzale Aldo Moro 5, 00185 Rome, Italy \\
         \and
             Departamento de Astronom\'{\i}a, Universidad de Chile, Casilla 36-D, Santiago, Chile.\\
             }

   \date{Received September xx, xxxx; accepted March xx, xxxx}

 
  \abstract
   {}
   {This study aims to elucidate the role of intermediate-mass black holes (IMBHs) in the development of galactic morphology. We will examine how the evolution of IMBHs is influenced by various factors, including seed masses, seed times, and feedback mechanisms. Additionally, we will investigate potential correlations between galactic morphology and the final properties of central dwarf galaxies, such as gas fraction, taking into account the constraints of black hole growth history and active galactic nucleus (AGN) feedback. This exploration will be particularly valuable in characterizing the typical environments associated with IMBHs in dwarf galaxies.}
   {We utilize a series of high-resolution zoom-in cosmological simulations to analyze the emergent morphology of central dwarf galaxies within dark matter halos of similar mass, specifically $\mathrm{{\rm log}\,M_{\rm vir}/{\rm M_\odot}\sim 10\pm 0.05}$, at redshift $\mathrm{z\sim 2}$. All simulations are initialized with the same conditions while employing different black hole seeding methods and feedback schemes. This approach enables us to investigate how varying black hole evolutionary pathways affect the fundamental parameters of central dwarf galaxies. Our simulations consider galaxies embedded in both high- and low-spin host halos, utilizing various black hole seed masses, seed formation times, and AGN wind velocities. }
  {This study investigates the impact of intermediate-mass black holes (IMBHs) on dwarf galaxy morphology at high redshift (z = 2) using cosmological simulations. We find that AGN feedback, particularly wind strength, critically influences gas fractions, star formation, and galaxy structure. Galaxies with strong feedback exhibit lower stellar masses, flatter morphologies, and intermediate rotational support ($\mathrm{\kappa_{rot}}$ = 0.3–0.6). Their prominent central structures and low Sersic indices ($\mathrm{n < 2}$) highlight the limits of applying low-redshift diagnostics like Gini-M20 at high redshift. Synthetic JWST observations suggest pixelation effects can overestimate galaxy sizes, emphasizing the nuanced relationship between IMBH evolution and dwarf galaxy formation.}
   {}

   \titlerunning{Feedback from intermediate mass black holes on dwarf galaxy morphology}
   \authorrunning{Bi, D., Schleicher, D. R, \& Escala A. }
   \maketitle
%

\section{Introduction}
\label{sec:intro}
Since the discovery of black holes (BHs), extensive discussions regarding their formation and growth mechanisms have emerged \citep[e.g.,][]{zel64,salpeter64,rees1984,ghez08,gill09}. The prevailing paradigm posits several formation pathways, including: (i) direct collapse \citep[e.g.,][]{bl03,lona06,lasch15,be15,redo18}, (ii) gas accretion \citep[e.g.,][]{mare01,vol03,ag13,rina18}, and (iii) catastrophic stellar collisions within dense stellar clusters \citep[e.g.,][]{zelpo65,rees1984,shte85,omu08,devo09,dev12,rei18,ver22}. Over the past few decades, black holes have been detected across a wide range of mass scales, from approximately $\mathrm{3M_\odot}$ \citep{tho19} to around $\mathrm{5 \times 10^{10} M_\odot}$ \citep{du21}. However, conclusive evidence for the existence of intermediate-mass black holes (IMBHs), defined as those with masses in the range of $\mathrm{10^3M_\odot \leq M_\bullet \leq 10^6M_\odot}$, remains elusive (see \citealp{greene04} for a review).

Fortunately, significant advancements in technology and observational instruments have facilitated an increasing number of studies that provide evidence supporting the existence of IMBHs. Ultraluminous X-ray emitters (ULXs) and hyperluminous X-ray sources (HLXs) are frequently proposed as hosts for IMBHs \citep{cm99}. The most prominent IMBH candidate currently identified is HLX-1, whose mass is estimated to lie between $\mathrm{3\times 10^3M_\odot}$ and $\mathrm{3\times 10^5M_\odot}$ (see the review by \citealp{mez17} and references therein). Other significant sources, such as NGC 5252 and NGC 2276-3C, are also estimated to harbor IMBHs with masses of approximately $\mathrm{10^5M_\odot}$ \citep{kim20} and $\mathrm{5\times 10^4M_\odot}$ \citep{mez13,mez15}, respectively. Additionally, the HLX M82 X-1 has also been suggested as a potential IMBH host \citep{psm14}.

Globular clusters represent another prime target for the search for IMBHs \citep{lah19}. For instance, the X-ray signal detected from G1 in M31 may indicate the presence of an IMBH with an estimated mass of about $\mathrm{2\times 10^4M_\odot}$ \citep{geb02,geb05}. Several studies have also reported the potential detection of a central black hole in $\omega$ Centauri, with the most recent work confirming a lower mass limit for this black hole of approximately $\mathrm{8,200 M_\odot}$, based on dynamical measurements of fast-moving stars in its vicinity \citep{hab24}. 

Dwarf galaxies also emerge as promising environments for the search for IMBHs. Based on the theoretical mechanisms outlined above, if the black hole seeding process occurred in relatively small halos at high redshifts, it remains possible for these halos to have persisted and to contain IMBHs within dwarf galaxies. Recent studies indicate that some supermassive black holes (SMBHs) in dwarf galaxies cross into the IMBH mass range \citep[e.g.,][]{gh07,bar08,rei13,bal15,bal18}. Numerous investigations, particularly at X-ray and infrared wavelengths, have been conducted to identify active galactic nuclei (AGN) in dwarf galaxies \citep[e.g.,][]{mez20,fmb21,mol21,bur22,srm22}. Furthermore, multiband observations facilitated by the upgraded LIGO detector and the proposed space mission LISA are expected to provide additional powerful tools for detecting IMBHs \citep{jani20}.

The understanding of the evolution of dwarf galaxies remains insufficiently refined. In recent years, numerous observational studies have been conducted to investigate these systems across cosmic time. However, discrepancies between observational results and theoretical predictions have led to several challenges. Notably, there is a mismatch between the number of observed Milky Way satellites and the predicted number of dark matter (DM) halos that may host these systems, a phenomenon known as the missing satellites problem \citep[e.g.,][]{kwg93,kly99,moo99}. Additionally, the absence of massive observed satellites poses another issue; specifically, the inferred central masses of the most massive observed satellites do not align with those of the most massive simulated subhaloes, leading to what is termed the too-big-to-fail problem \citep[e.g.,][]{bbk11}. Furthermore, there exists a discrepancy between the inferred DM halo profiles of observed dwarfs and the cuspy DM profile predicted by the Lambda Cold Dark Matter (LCDM) model, referred to as the cusp versus core problem \citep[e.g.,][]{fp94,moo94}. Collectively, these issues have generated significant interest within the astronomical community.

Moreover, akin to the role of supermassive black holes (SMBHs), the presence of intermediate-mass black holes (IMBHs) in dwarf galaxies is suggested to provide active galactic nucleus (AGN) feedback. Due to the shallow potential wells of dwarf galaxies, it is posited that such feedback may play a more critical role in regulating the baryon cycle, making dwarf galaxies ideal locations to investigate how black holes influence galaxy evolution \citep[e.g.,][]{cr22}. 

Observational studies have confirmed that IMBHs are challenging to detect at high redshift, particularly beyond $\mathrm{z > 7}$ \citep[e.g.,][]{sob15,ag16,pac16,smith16,pez17}. However, it has been posited that the nuclei of dwarf galaxies could serve as optimal sites for the search for IMBHs at lower redshifts, representing the remnants of seed black holes from higher redshifts (see review by \citealp[]{mez17}). Several attempts have been made to locate IMBHs in dwarf galaxies, employing various methods such as X-ray emission detection \citep[e.g.,][]{gh07,des09,rei11,dong12,sch13,bal15,bal17,lem15,sec15,par16,chen17,mez18}, standard virial techniques to estimate black hole masses \citep[e.g.,][]{bar04,gh04,gh07,pet05,rei13,lf15,ben16,non17,liu18}, and searches in the infrared (IR) regime \citep[e.g.,][]{sa07,sa08,sa09,sa14,sar15,mar17}. However, it is important to note that most of these observational efforts may be biased due to factors such as sample selection at specific redshifts (typically $\mathrm{z < 0.3}$), as well as considerations regarding black hole Eddington ratios and AGN types.

In light of the limitations posed by observational challenges, numerical simulations have emerged as indispensable tools for testing galaxy formation and elucidating the relationships between black holes and their host galaxies, particularly at high redshift. Over the past few decades, numerous numerical attempts have focused on black hole models in dwarf galaxies. These include large box cosmology simulations \citep[e.g.,][]{vogel14,nel18,nel19,schaye15}, which provide ample sample volumes to trace the evolution of scaling relations. However, such simulations often lack sufficient mass and spatial resolution at the lower mass end where dwarf galaxies reside. In contrast, high-resolution simulations in smaller volumes have concentrated on black holes within dwarf galaxies \citep[e.g.,][]{bar19,kss22,lan21}. These studies typically evaluate the growth and feedback of black holes and explore the correlation between black hole growth and star formation quenching in host galaxies. Nevertheless, most analyses frequently overlook the morphology of the host galaxy and do not exclude potential influences related to the host dark matter halo, such as halo spin. Indeed, statistical findings from numerical simulations indicate that halo spin plays a significant role for dwarf galaxies \citep[]{rodr17}.

Moreover, it is widely accepted that the morphological characteristics of galaxies in the contemporary universe result from the convergence of numerous physical processes—many of which remain inadequately understood and quantified \citep[e.g.,][]{shlo13}. These processes include the impacts of AGN feedback, environmental factors (encompassing parent dark matter halos), stellar feedback, local and global instabilities within stellar and gaseous disks, as well as the origins of stellar bulges and bars \citep[e.g.,][]{hell07}.

Considering the properties of dwarf galaxies outlined earlier, this study concentrates on exploring the correlation between dwarf galactic morphologies and active galactic nucleus (AGN) feedback from intermediate-mass black holes (IMBHs) within cosmological simulations at high redshift ($\mathrm{z \gtrsim 2}$). By employing the zoom-in technique, we achieve high resolution and quantify galactic morphology through various methods, including nonparametric tools such as the Gini-$\mathrm{M_{20}}$ method and the Concentration-Asymmetry-Smoothness (CAS) method \citep{con03,lotz04}, two-dimensional Sersic profile fitting, and stellar kinematics analysis. We vary the AGN feedback parameters within the same galaxy and compare the resultant outputs to assess how AGN feedback influences our numerical models.

AGN feedback primarily manifests as wind feedback; both simulation and observational studies indicate that it is directly linked to quenching star formation (SF) \citep[e.g.,][]{ssb05,bar14,sch06,lanz16}. Conversely, AGN outflows may also compress clumpy gas clouds, potentially triggering starbursts and thereby facilitating positive feedback \citep[e.g.,][]{dy89,bieri15,mu18,cmb87,vg92,zinn13,sal17}. The effects of AGN feedback are contingent upon numerous factors, including black hole (BH) mass, accretion rate, and feedback intensity. Consequently, we evaluate the possible impacts within the same galaxy under different combinations of BH seeding time, mass, and feedback intensities. Our aim is to elucidate the role that these IMBHs play in the evolution of galactic morphology, particularly by investigating potential correlations between morphology and the final properties of the central dwarf galaxy, such as gas fraction, while considering the constrained history of BH growth and AGN feedback. This research is particularly valuable for interpreting future observations of IMBHs and understanding how representative the observed masses may be in relation to their original masses at formation.

Given the limitations of our computational resources, our selection of models is informed by several criteria: (1) focusing on galaxies residing in halos with a mass range of approximately $\mathrm{10^{10} M_\odot}$, where dwarf galaxies are anticipated to form at the center; (2) introducing different BH seedings and feedbacks within the same target galaxy from very high redshift, as these effects are expected to influence the outcome at the final redshift ($\mathrm{z \sim 2}$ in our case); and (3) randomly selecting galaxies from two groups based on the spins of their host dark matter (DM) halos. This approach enables us to assess the potential impacts arising from the bias of the host DM halo on galaxy morphology.

The structure of this paper is organized as follows: Section 2 outlines the numerical methods and issues encountered. Section 3 presents our results, encompassing general properties, morphology, and kinematic analysis of the galaxies. This is followed by a discussion section and a summary.

\section{Numerical modeling}
\label{sec:modeling}

\subsection{Numerics and initial conditions}
\label{sec:numerics}
We utilized a modified version of the hybrid $\mathrm{N}$-body/hydro code \textsc{gizmo} \citep{hopk17}, which features the Lagrangian meshless finite mass (MFM) hydrodynamics solver. \textsc{gizmo} incorporates an enhanced version of the TreePM gravity solver originally developed for \textsc{gadget} \citep{spri05}. 

We adopt the Planck\,16 $\mathrm{\Lambda}$CDM concordant cosmology with parameters $\mathrm{\Omega_{\rm m} = 0.308}$, $\mathrm{\Omega_\Lambda = 0.692}$, $\mathrm{\Omega_{\rm b} = 0.048}$, $\mathrm{\sigma_8 = 0.82}$, and $\mathrm{n_{\rm s} = 0.97}$ \citep{planck16}. The Hubble constant is defined as $\mathrm{h = 0.678}$ in units of $\mathrm{100\,{\rm km\,s^{-1}\,Mpc^{-1}}}$. The initial conditions were established at a starting redshift of $\mathrm{z=99}$. Both the parent box and the individual zoom-in simulations were generated using the code \textsc{music} \citep{hahn11}. The simulations progressed from $z = 99$ to the target redshift $\mathrm{z_{\rm f} = 2}$.

The MFM hydrosolver employs adaptive gravitational softening for the gas. An explicit mechanical algorithm is utilized for Supernova feedback \citep{hopk18}. The simulations also incorporate a redshift-dependent cosmic ultraviolet (UV) background \citep{Faucher09}.

We utilize the astrochemistry package KROME (\citealp[]{gra14}) for modeling gas cooling and heating processes. The model is consistent with that employed by Bovino et al. (2016), which incorporates several mechanisms, including photoheating, H2 UV pumping, Compton cooling, photoelectric heating, atomic cooling, H2 cooling, and both chemical heating and cooling.

In the zoom-in region of our study, we operate with a mass per particle of $\mathrm{3.5 \times 10^4\,{\rm M_\odot}}$ for gas and stars, and $2.3 \times 10^5\,{\rm M_\odot}$ for dark matter (DM). In comoving coordinates, the minimum adaptive gravitational softening length is 74 pc for gas, while it is 74 pc, 7.4 pc, and 118 pc for stars, DM, and black holes (BHs), respectively. The effective number of baryonic particles in our simulations totals $\mathrm{2 \times 4096^3}$.

\subsection{Star formation}
\label{sec:starformation}

Star formation is modeled using a turbulence-driven approach as described by \citealp[]{lupi18}. This methodology employs a stochastic prescription that converts gas particles into stellar particles, with the star formation rate (SFR) density defined by the relation:
\begin{equation}
\dot{\rho}_{SF}=\epsilon \frac{\rho_{g}}{t_{ff}}
\end{equation}
Here, $\mathrm{\epsilon}$ represents the star formation efficiency parameter, $\mathrm{t_{ff}=\sqrt{\frac{3 \pi}{32 G \rho_{g}}}}$ denotes the free-fall time, $\rho_{g}$ is the local gas density, and $\mathrm{G}$ is the gravitational constant. The efficiency parameter is expressed as:
\begin{equation}
\epsilon=\frac{\epsilon_\star}{2\phi_{\rm t}}\exp\left({\frac{3}{8}\sigma_s^2}\right)\left[1+{\rm erf}\left({\frac{\sigma_s^2-s_{\rm crit}}{\sqrt{2\sigma_s^2}}}\right)\right],
\end{equation} 
where $\mathrm{\epsilon_\star=0.5}$ is the local star formation efficiency adjusted to match observational data (Heiderman et al. 2010), and $\mathrm{1/\phi_{\rm t}=0.49 \pm 0.06}$ is a calibration factor that accounts for uncertainties in the free-fall time scale. The parameter $\mathrm{s_{\rm crit}}$ is defined as $\mathrm{s_{\rm crit}= \ln{(\rho_{\rm crit}/\rho_0)}}$.

Star formation is permitted only when the gas satisfies two criteria: (i) $\mathrm{\rho_{g} > \rho_{SF}}$ and (ii) the Mach number $\mathrm{\mathcal{M} > 2}$.

\subsection{Black hole physics.}
\label{sec:bhphysics}
To investigate various sub-resolution models for black holes (BHs), we conduct a series of simulations that vary three key parameters: the initial seed BH mass ($\mathrm{M_{seed}}$), the minimum halo mass for BH seeding ($\mathrm{M_{HaloMin}}$), and the AGN wind velocity for BH feedback ($\mathrm{v_{wind}}$). These parameters have been observed to significantly influence BH growth and feedback mechanisms.

In our simulations, the seeding of BHs relies on global information regarding the host halos, which is provided by an on-the-fly group finder operating during the simulation. BHs are initiated at the center of mass of the host halo once it reaches a predetermined mass threshold. Due to the limitations of our simulation in resolving traditional accretion disk radii, we employ a "sub-grid" torques-driven accretion model (\citealp[]{hq11}) wherein the accretion rate is formulated as follows:
\begin{equation}
\dot{M}_{BHL}\sim \alpha \epsilon_{acc} \frac{M_{gas}(<R)f^{1/2}_dM^{1/6}_{BH,8}R^{-2/3}_{100}}{2\times10^8 \, \mathrm{yr} \,(f_{gas}+0.3f_{d}M^{-1/3}_{d,9})} ,
\end{equation}
In this equation, $\mathrm{\alpha}$ is an arbitrary factor that amplifies the calculated accretion rate, $M_{BH,8}\equiv M_{BH}/10^8M_\odot$, $M_{d,9}\equiv M_{disk}/10^9M_\odot$, $f_{d}\equiv M_{disk}(<R)/M_{total}(<R)$, $f_{gas}\equiv M_{gas}(<R)/M_{total}(<R)$, and $R_{100}\equiv R/100 \, \mathrm{pc}$. Here, $R$ is defined as three times the full extent of the force softening kernel.\\

In our simulations, we set the accretion boost factor to $\mathrm{\alpha = 1}$. The fundamental properties of AGN wind feedback are typically characterized by two parameters: the mass loading $\mathrm{\beta = \dot{M_{wind}}/\dot{M_{BH}}}$ and the wind velocity $\mathrm{v_{wind}}$. In the AGN wind model employed in our study, mass, energy, and momentum are continuously injected into the gas surrounding the black hole (BH) (within its neighboring kernel) at a rate that corresponds to the specified accretion rate and mass-loading. Fractional weights are utilized to allocate energy and momentum to neighboring gas elements on a per-timestep basis (\citealp[]{hopk18a}). In accordance with observations of AGN ultra-fast outflows (e.g., \citealp[]{mg15}; \citealp[]{ktb18}, and references therein), we have set our mass loading to $\mathrm{\beta = 1}$ and the wind velocity to $\mathrm{v_{wind} = 10,000 \, \mathrm{km\,s^{-1}}}$. For comparison, we also incorporate models with the same parameter settings but with a relatively smaller wind velocity of $\mathrm{v_{wind} = 2,000 \, \mathrm{km\,s^{-1}}}$.

Due to resolution limitations, the dynamical friction estimator may not perform effectively, potentially leading to artificial behaviors of BHs caused by particle noise and local clumps. To mitigate this issue, we compel the BH to move to a local potential minimum within the kernel. This approach considers only the neighboring star particles $\mathrm{b}$ for positioning and restricts movement to those positions where the relative velocity $\mathrm{\left|v_b - v_a \right|^2 < v^{2}_{esc, BH} + c^{2}_{s, BH}}$, with $\mathrm{c_s, BH}$ representing the sound speed of the gas in the vicinity of the BH and $\mathrm{v_{esc, BH}}$ denoting the escape velocity from the BH at a distance $\mathrm{\left|x_b - x_a \right|}$. This method ensures that the BH can move smoothly, avoiding large positional jumps.

We do not implement a “sub-grid” model for recoils or ejections in BH-BH mergers. Whenever two BHs are within the same smoothing kernel or at the resolution limit, we merge them if their relative velocities fall below the mutual escape velocity of the two-BH system at the resolved separation.

In summary, we vary the wind velocity parameter $\mathrm{v_{wind}}$ to adjust the momentum-loading of the winds, thereby controlling the feedback efficiency. Additionally, we modify the minimum host dark matter (DM) halo mass $\mathrm{M_{vir}}$ to regulate the BH seeding time and directly adjust the seed BH initial mass $\mathrm{M_{seed}}$. The simulation models are named in the format BHs[X]hl[Y]v[Z][L/H], where X, Y, and Z represent the BH initial mass $\mathrm{M_{seed}}$, the minimum host DM halo mass $\mathrm{M_{vir}}$ criteria, and the AGN wind velocity $\mathrm{v_{wind}}$, respectively. The last suffix, L/H, indicates whether the model corresponds to a galaxy in a DM halo with low or high spin. Models without BHs are labeled as NoBH[L/H]. Comprehensive details regarding the parameter sets used in our simulations are presented in Table \,\ref{tab:BHsim}.

\begin{table*}
\caption{Overview of the parameter set of cosmological zoom-in simulation runs. The simulations with galaxies in low (upper) and high (lower) spin halos are listed separately. The details of our model naming rules can be found in Section\,\ref{sec:bhphysics}. We list the host DM halo mass ($\mathrm{M_{vir}}$), spin ($\mathrm{\lambda}$), AGN wind velocity ($\mathrm{v_{wind}}$), BH initial mass ($\mathrm{M_{seed}}$) and minimum halo mass for BH seeding ($M_{HaloMin}$). The gravitational/torque/angular-momentum-driven BH accretion model \citep[]{hq11} has been used. Following the observations and theoretical models suggestion, we set all AGN wind models with mass loading $\mathrm{\beta=1}$. } 
\label{tab:BHsim}
\vspace{-.1cm}
\centering
\begin{tabular}{ccccccccccc}
\hline
\hline
 Model No. &$\mathrm{z_{\rm f}}$ & $\mathrm{{\rm log}\,M_{\rm vir}}\,$& $\mathrm{\lambda}$&  $\mathrm{v_{wind}}$ & $\mathrm{M_{seed}}$ & $\mathrm{M_{HaloMin}}$ & BH present \\
         
  & & [M$_\odot$] &   & [$\mathrm{km\,s^{-1}}$] & [M$_\odot$] & [M$_\odot$] &  \\
\hline
    NoBHL     & 2 & $\sim 10$ & $\sim0.035$&  N/A & N/A& N/A & No  \\
\hline
  
    BHs4hl6v2L    & 2 & $\sim 10$ & $\sim0.035$& 2000 & $5\times 10^4$& $10^6$ & Yes  \\

\hline   
    BHs4hl7v2L    & 2 &  $\sim 10$&$\sim0.035$&  2000 & $5\times 10^4$& $10^7$  & Yes  \\
\hline 
    BHs4hl6v10L    & 2 & $\sim 10$ & $\sim0.035$&  10000 & $5\times 10^4$& $10^6$ & Yes  \\

\hline   
   BHs4hl7v10L    & 2 &  $\sim 10$&$\sim0.035$&  10000 & $5\times 10^4$& $10^7$  & Yes  \\
\hline 
  
    BHs3hl6v2L     & 2 & $\sim 10$ & $\sim0.035$&  2000 & $5\times 10^3$& $10^6$ & Yes  \\

\hline   
    BHs3hl7v2L   & 2 &  $\sim 10$&$\sim0.035$&  2000 & $5\times 10^3$& $10^7$  & Yes  \\
\hline 
   BHs3hl6v10L     & 2 & $\sim 10$ & $\sim0.035$&  10000 & $5\times 10^3$& $10^6$ & Yes  \\

\hline   
    BHs3hl7v10L     & 2 &  $\sim 10$&$\sim0.035$&  10000 & $5\times 10^3$& $10^7$  & Yes  \\
\hline 
\hline
    NoBHH    & 2 & $\sim 10$ & $\sim0.1$& N/A &  N/A& N/A & No  \\
\hline
  
    BHs4hl6v2H     & 2 & $\sim 10$ & $\sim0.1$&  2000 & $5\times 10^4$& $10^6$ & Yes  \\

\hline   
    BHs4hl7v2H   & 2 &  $\sim 10$&$\sim0.1$&  2000 & $5\times 10^4$& $10^7$  & Yes  \\
\hline 
    BHs4hl6v10H   & 2 & $\sim 10$ & $\sim0.1$& 10000 & $5\times 10^4$& $10^6$ & Yes  \\

\hline   
    BHs4hl7v10H    & 2 &  $\sim 10$&$\sim0.1$&  10000 & $5\times 10^4$& $10^7$  & Yes  \\
\hline 
  
    BHs3hl6v2H     & 2 & $\sim 10$ & $\sim0.1$& 2000 & $5\times 10^3$& $10^6$ & Yes  \\

\hline   
    BHs3hl7v2H    & 2 &  $\sim 10$&$\sim0.1$&  2000 & $5\times 10^3$& $10^7$  & Yes  \\
\hline 
   BHs3hl6v10H     & 2 & $\sim 10$ & $\sim0.1$&  10000 & $5\times 10^3$& $10^6$ & Yes  \\

\hline   
   BHs3hl7v10H     & 2 &  $\sim 10$&$\sim0.1$&  10000 & $5\times 10^3$& $10^7$  & Yes  \\
\hline   
\hline
\end{tabular}
\end{table*}

\subsection{Halo and galaxy finder}
\label{sec:group}
The \textsc{music} algorithm \citep{hahn11} has been utilized for establishing the initial conditions for both the parent box and individual zoom-in simulations. \textsc{music} employs a real-space convolution approach along with an adaptive multi-grid Poisson solver to construct highly accurate nested density, particle displacement, and velocity fields, which are suitable for multi-scale zoom-in simulations of cosmic structure formation. Initially, a comoving box of 50 Mpc was created with $\mathrm{1024^3}$ dark matter (DM) particles, which was then run until a redshift of $\mathrm{z=2}$. DM halos were selected by applying a group finder at the target redshift of $\mathrm{z=2}$, to attain the prescribed DM masses of log\,($\mathrm{M_{\rm vir}/{\rm M_\odot}) \sim 10.0 \pm 0.05}$ and dimensionless halo spin $\mathrm{\lambda}$ (see below for definitions of $\mathrm{M_{\rm vir}}$ and $\mathrm{\lambda}$).

The DM halos were identified using the \textsc{rockstar} group finder \citep{behr13}, with a Friends-of-Friends (\textsc{FoF}) linking length of $\mathrm{b=0.28}$ and the removal of unbound DM particles. The halo virial radius and the virial mass, denoted as $\mathrm{R_{\rm vir}}$ and $\mathrm{M_{\rm vir}}$, were calculated based on the definitions of $\mathrm{R_{200}}$ and $\mathrm{M_{200}}$ \citep[e.g.,][]{nfw96}. Here, $\mathrm{R_{200}}$ represents the radius within which the mean interior density is 200 times the critical density of the universe at that time. The dimensionless halo spin, $\mathrm{\lambda}$, is defined as \citep[e.g.,][]{bull01}:
\begin{equation}
\lambda = J/J_{\rm max},    
\end{equation}
where $\mathrm{J}$ denotes the DM halo angular momentum, and $\mathrm{J_{\rm max}}$ signifies the maximal (Keplerian) angular momentum. We randomly selected two halos with similar mass ranges, specifically $\mathrm{M_{\rm vir} \sim 10^{10} M_\odot}$: one halo has an average spin $\mathrm{\sim 0.035}$, while the other is characterized by a relatively high spin of $\mathrm{\sim 0.1}$.

Subsequently, we carved out a sphere encompassing all particles within a volume of $\mathrm{\sim 5R_{\rm vir}}$ to mitigate contamination of the high-resolution region by massive particles. These particles were traced back to their initial conditions to create a mask for \textsc{music}, where the resolution was enhanced by 5 levels, transitioning from $\mathrm{2^8}$ to $\mathrm{2^{12}}$. Ultimately, we utilized 4 levels of refinement, focusing exclusively on the highest refinement level.

We applied the group-finding algorithm \textsc{hop} \citep{eise98} to identify galaxies, using an outer boundary threshold of baryonic density $\mathrm{0.04\,\text{cm}^{-3}}$. This density threshold ensures that both the host star-forming gas and the lower-density, non-star-forming gas are approximately bound to the galaxy \citep{roma14}. We employed \textsc{hop} for this purpose to avoid imposing a specific geometric shape on the galaxy. It is important to note that this definition of a galaxy differs from many used in the literature, which often base galaxy size on a fraction of $\mathrm{0.1R_{\rm vir}}$ \citep[e.g.,][]{mari14}.

\subsection{From stellar particles to galaxy light: the synthetic imaging}
\label{sec:light}

To ascertain the observational properties of our galaxies, we employed the 3-D radiation transfer code \textsc{SKIRT} \citep{baes03} for post-processing. This involved redshifting and pixelizing the data, ultimately producing synthetic images of the galaxies as viewed face-on. Each stellar particle was treated as a coeval single stellar population characterized by the initial mass function (IMF) from \citep{chab03}, with an associated spectral energy distribution (SED) family derived from Bruzual and Charlot \citep{bruz93}. The density of dust was assumed to be proportional to the gas metallicity, with a proportionality constant of 0.3 \citep[e.g.,][]{camps16}. The total emission accounted for not only the direct emission from the stars but also the secondary emissions due to dust, including multiple iterations for dust self-absorption. For our analysis, we selected the F200W filter from the James Webb Space Telescope (JWST) to achieve the brightest observational images (see Section\,\ref{sec:photo}).

\section{Results}
\label{sec:results}
Our analysis is based on two randomly selected galaxies, one residing in a low-spin dark matter (DM) halo and the other in a high-spin halo. Each galaxy encompasses a range of parameters, including the initial mass of black hole (BH) seeding, seeding time, and AGN wind velocity. Some results may be influenced by the specific properties of the selected galaxies. However, while we do not claim statistical significance, this study serves as a valuable entry point for examining how intermediate-mass black holes (IMBHs) influence the morphology of galaxies across specific mass ranges, such as dwarf galaxies.

It is important to note that our selection of zoom-in target galaxies means that these dwarf galaxies are typically the central galaxies within their host DM halos. As a result, the evolutionary trajectory of the central galaxy may substantially differ from that of satellite galaxies \citep[e.g.,][]{wang20}.

To elucidate the evolutionary details, we commence by presenting the evolution of general parameters of the galaxies. Models without BHs serve as benchmarks, enabling comparison of the effects of BHs on galaxy evolution. For clarity, we focus on the last 1.8 billion years of our models, spanning redshifts from $\mathrm{z=4}$ to $\mathrm{z=2}$.

Various methodologies exist for quantifying galactic morphology from imaging data. In this study, we assess both parametric and non-parametric approaches by modeling the surface intensity of the galaxy images. We employed the galaxy morphology tool \textsc{STATMORPH} \citep{rodr19} to extract the Gini coefficient, $\mathrm{M_{20}}$, as well as C, A, and S measurements for each galaxy. According to \citealp[]{tho15}, the Gini-$\mathrm{M_{20}}$ method is less effective for edge-on or randomly oriented galaxies than for face-on systems. Therefore, all two-dimensional analyses in this work are conducted exclusively on face-on images, which were generated by orienting the galaxies according to the total angular momentum of the stellar particles. Additionally, we applied Sersic profile fittings to the same images. Kinematic energy parameters, denoted as $\mathrm{\kappa_{rot}}$, were also calculated to evaluate the rotational support of each galaxy. Ultimately, we repeated the morphological measurements for the photometric images of our face-on galaxies at the final redshift.

For clarity and conciseness, we present the results of our analysis separately for host halos exhibiting low and high spin. Based on the parameters set for black holes (BHs), we can effectively categorize the models into distinct groups: (i) BHs3hl6v2[L/H] represents the model that seeds a black hole with a lower mass ($\mathrm{10^{3} M_\odot}$) at an earlier stage (when the host halo reaches $\mathrm{10^{6} M_\odot}$); (ii) BHs3hl7v2[L/H] refers to the model that seeds a black hole with a lower mass ($\mathrm{10^{3} M_\odot}$) at a later stage (when the host halo reaches $\mathrm{10^{7} M_\odot}$); (iii) BHs4hl6v2[L/H] denotes the model that seeds a black hole with a higher mass ($\mathrm{10^{4} M_\odot}$) at an earlier stage (when the host halo reaches $\mathrm{10^{6} M_\odot}$); and (iv) BHs4hl7v2[L/H] represents the model that seeds a black hole with a higher mass ($\mathrm{10^{4} M_\odot}$) at a later stage (when the host halo reaches $\mathrm{10^{7} M_\odot}$). A similar categorization applies to models demonstrating stronger BH feedbacks, characterized by an AGN wind velocity of approximately $\mathrm{10000\, km\,s^{-1}}$. It is important to note that the data for the model BHs4hl5v10H appears to be truncated around redshift $\mathrm{z \sim 3}$. This truncation arises because this particular galaxy is insufficiently dense to be classified as a galaxy until approximately $\mathrm{z \sim 3}$, according to the definitions employed in our galaxy-finding algorithm.

\subsection{Galaxies: general}
\label{sec:galaxies}

\begin{figure*}
 \includegraphics[width=0.9\textwidth]{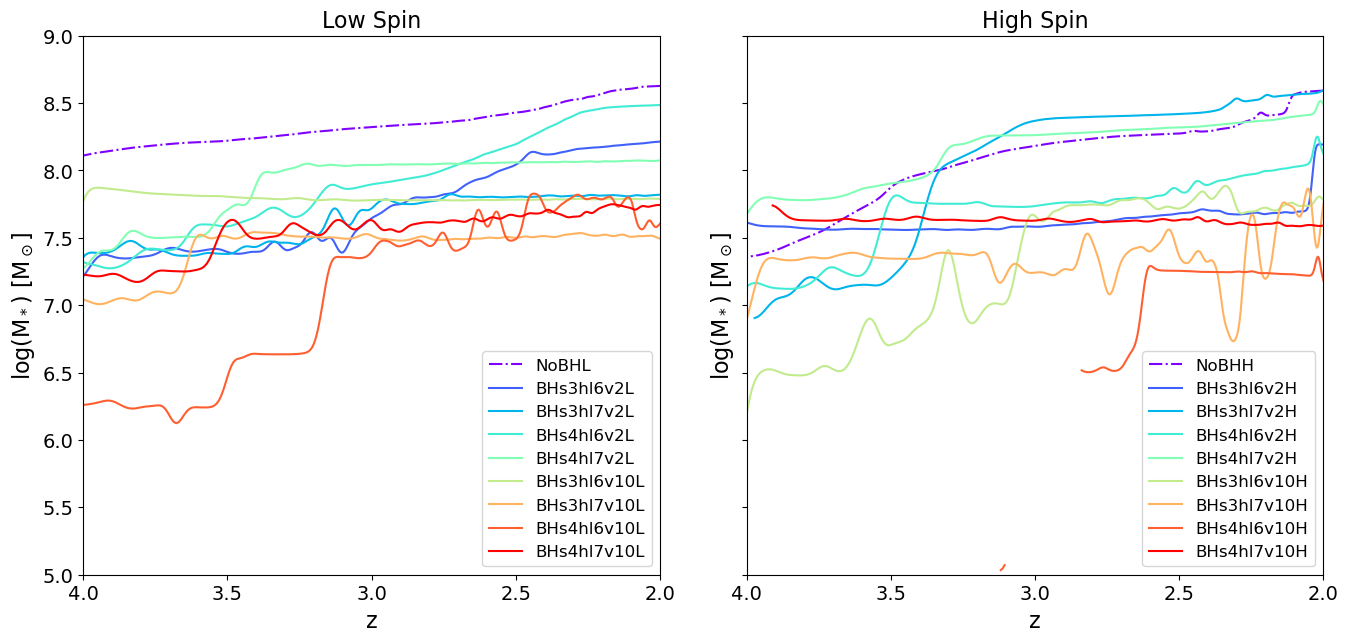}
    \caption{Evolution of stellar masses for galaxies in low spin (left) and high spin (right) host halos, from redshift $\mathrm{z=4}$ to 2. All models in Table\,\ref{tab:BHsim} are labeled with different colors.}
    \label{fig:stellar_mass}
    \end{figure*} 

\begin{figure}
 \includegraphics[width=0.49\textwidth]{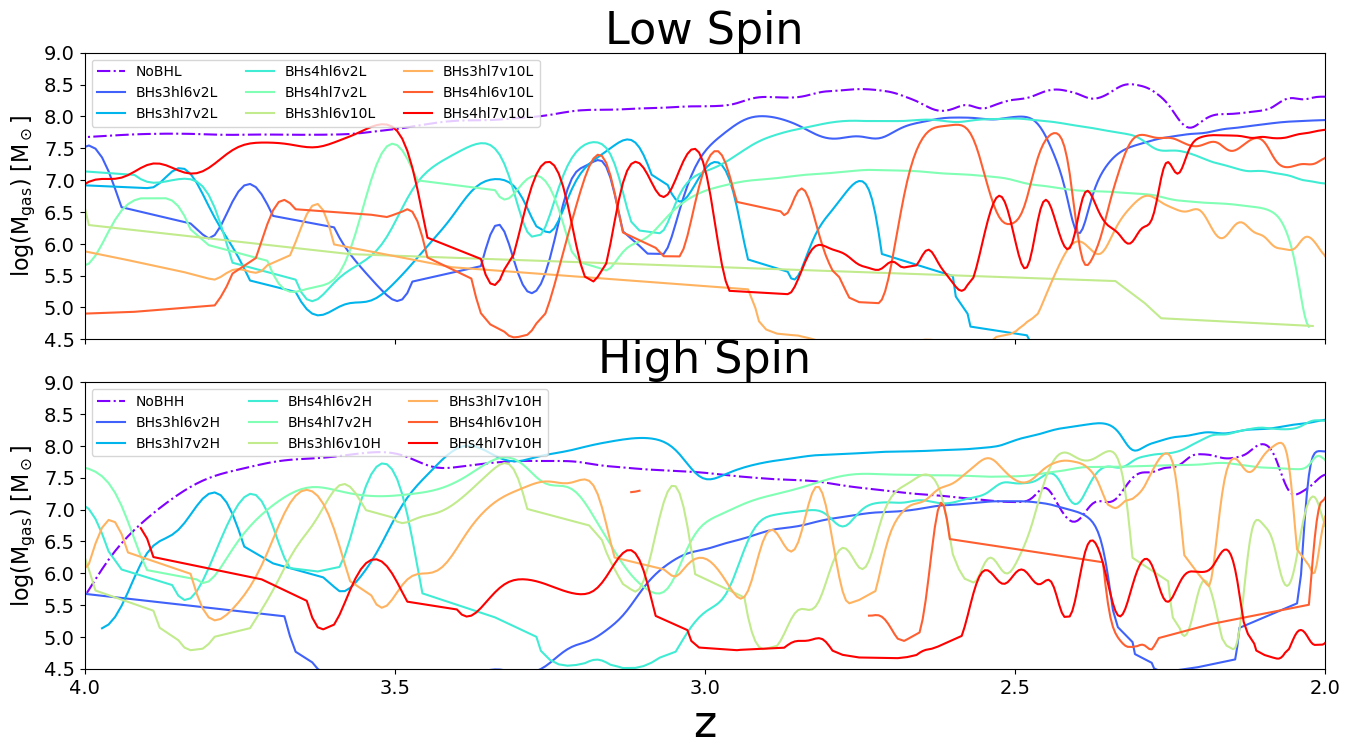}
    \caption{Evolution of gas mass ($\mathrm{M_{gas}}$) in galaxies in low spin (upper) and high spin (lower) host halos, from redshift $\mathrm{z=4}$ to 2. All models in Table\,\ref{tab:BHsim} are labeled with different colors. }
    \label{fig:gas_frac}
    \end{figure} 

Figure\,\ref{fig:stellar_mass} illustrates the evolution of stellar masses in galaxies\footnote{In Model BHs3hl7v10H, the central galaxy is insufficiently massive to be identified by the galaxy finder prior to $\mathrm{z\sim3}$.}. The models with black holes (BHs) are represented by solid lines, while those without BHs are depicted as dashed lines, each in distinct colors. Given that the host dark matter (DM) halos are configured within a mass range of approximately $\mathrm{10^{10} M_\odot}$, the final stellar mass of our central galaxies fluctuates between $\mathrm{10^{7.5}M_\odot \ltorder M_\star \ltorder 10^{8.5} M_\odot}$ at redshift $\mathrm{z=2}$. The final stellar mass of galaxies exhibits no significant dependence on the halo spin parameter $\mathrm{\lambda}$, yet shows a strong correlation with the BH models. 

The stellar mass growth in models without BHs (NoBHL/H) and in models with BHs characterized by low AGN wind velocities (e.g., BHs3hl6v2L, BHs4hl7v2L/H, HBHs4hl6v2L/H, BHs4hl7v2L/H) tends to be monotonically increasing, culminating in relatively higher stellar masses, with the exception of model BHs3hl6v2H. Conversely, models exhibiting high AGN wind velocities typically display flattened growth curves after $\mathrm{z\sim3.5}$, concluding with comparatively lower stellar masses (e.g., BHs3hl6v10H/L, BHs4hl6v10H/L, BHs4hl7v10H/L). Notably, models with higher initial BH masses, earlier seeding times, and high AGN wind velocities (e.g., BHs4hl6v10L/H) manifest growth patterns resembling step functions, suggesting that a significant portion of stellar mass increase in these galaxies arises from mergers. 

In addition to differences in growth slopes, we observe that the growth curves of models without BHs or with low AGN wind velocities exhibit fewer noise peaks compared to those with high AGN wind velocities. These noise peaks are typically associated with minor mergers and other catastrophic tidal events, which we will examine further in subsequent analyses. Figure\,\ref{fig:gas_frac} presents the evolution of gas masses across all models. Despite some fluctuations, galaxies devoid of BHs consistently maintain gas fractions ($\mathrm{M_{gas}/M_*}$) exceeding 10\% throughout the interval from $\mathrm{2 \ltorder z \ltorder 4}$. Conversely, galaxies containing BHs experience significantly more dramatic variations in gas mass—the gas fraction can swiftly drop to 0\% and subsequently recover to its original level. Such processes tend to recur more frequently in models characterized by elevated AGN wind velocities. In some extreme cases, such as model BHs3hl6v2H, the galaxy becomes nearly dry from $\mathrm{z=4}$, remaining dry (with a gas fraction lower than 10\%) until the final redshift $\mathrm{z=2}$. This phenomenon also accounts for the flat stellar mass curves discussed previously. In the absence of gas, the star formation rates in these galaxies decline sharply, as we will illustrate next. 

\begin{figure}
 \includegraphics[width=0.49\textwidth]{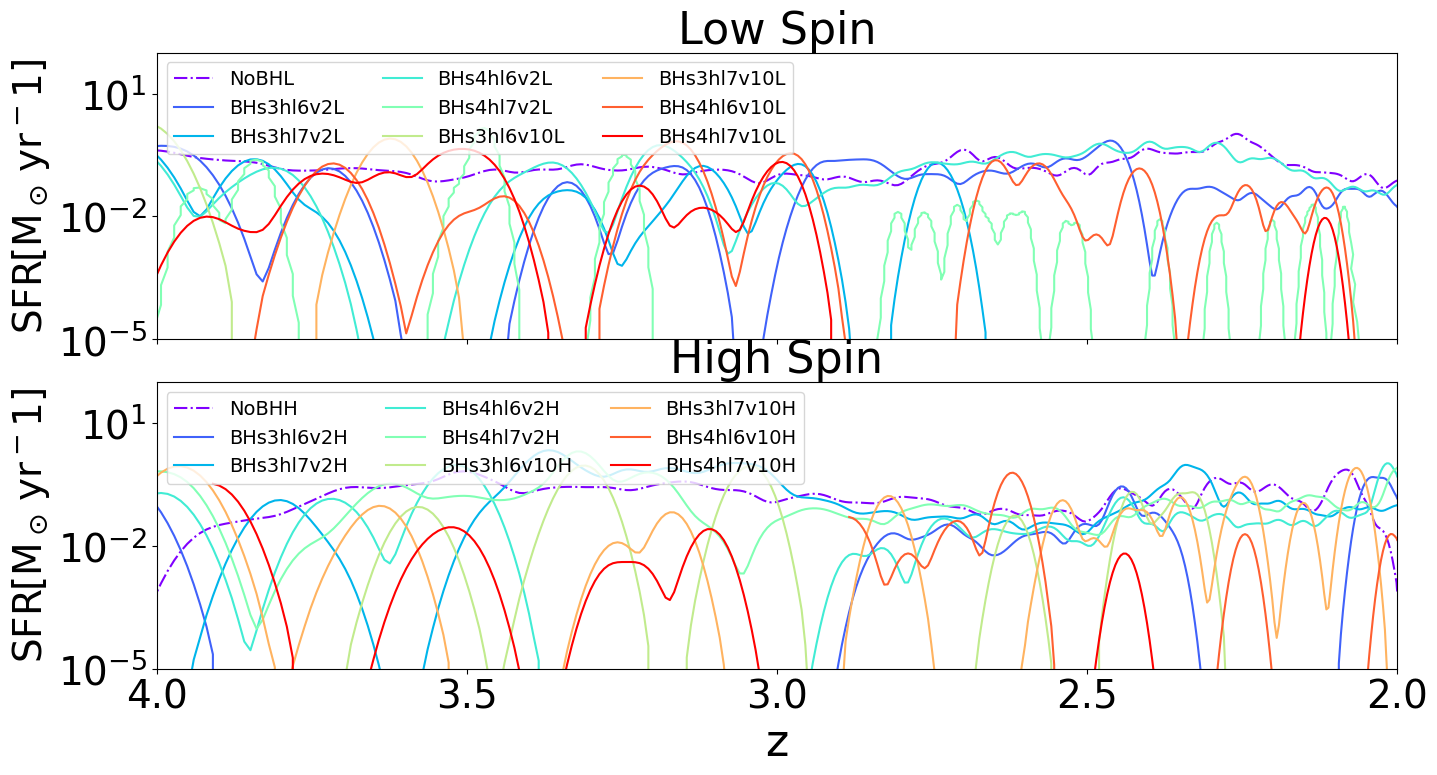}
    \caption{Evolution of SFR in galaxies in low spin (left) and high spin (right) host halos, from redshift $\mathrm{z=4}$ to 2. All models in Table\,\ref{tab:BHsim} are labeled with different colors. }
    \label{fig:sfr}
    \end{figure} 

\begin{figure}
 \includegraphics[width=0.48\textwidth]{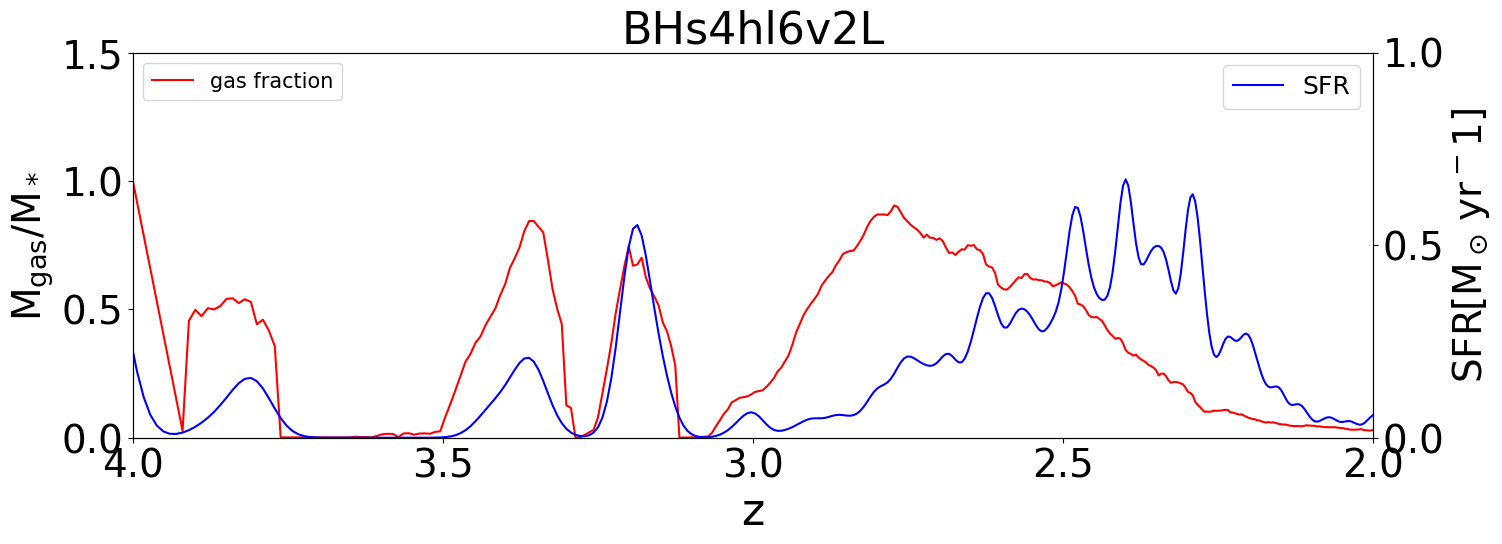}
    \caption{Correlation between gas fraction (red) and SFR (blue) for representative model BHs4hl6v2L, from redshift $\mathrm{z=4}$ to 2. }
    \label{fig:gasfrac_vs_sfr}
    \end{figure} 
    
Figure \,\ref{fig:sfr} presents the star formation rate (SFR) of galaxies as a function of redshift. It is evident that models without black holes (BHs) exhibit considerably more stable and continuous SFRs, averaging around $\mathrm{\sim 0.25 - 0.3 \, M_\odot\, yr^{-1}}$, interspersed with a few peaks in SFR (specifically, NoBHL at $\mathrm{z = 2.3}$ and NoBHH at $\mathrm{z = 3.5}$ and $\mathrm{z = 2.1}$). In contrast, models incorporating BHs experience frequent starbursts, a phenomenon associated with galaxy quenching. The amplitude of the SFR can increase by more than a factor of ten relative to the past averaged SFR, particularly for galaxies situated in high spin halos. When compared to the previously illustrated variability of the gas fraction (depicted in Figure\,\ref{fig:gas_frac}), a positive correlation between gas fraction and SFR is readily discernible. The model BHs4hl6v2L serves as a pertinent example of this correlation. In Figure\,\ref{fig:gasfrac_vs_sfr}, we plot the evolutionary trajectories of both SFR (blue) and gas fraction (red). Notable fluctuations in the gas fraction curve occur during the redshift ranges $\mathrm{z = 3.9 - 3.7}$, $\mathrm{z = 3.5 - 3.1}$, and $\mathrm{z = 3.1 - 2.0}$. The SFR exhibits a similar pattern, with each peak in gas fraction corresponding to a bump in SFR, albeit sometimes with temporal delays. This correlation is consistently observed, even among models featuring varying BH-related parameters (such as seeding initial mass, time, or wind velocity), indicating that in these galaxies, the upper limit of SFR is primarily constrained by the available gas reservoir. In many cases, star formation proceeds so efficiently that it can rapidly deplete the galaxy's gas supply. Subsequently, star formation ceases until new gas accrues in the galaxy. This cycle may repeat several times throughout the galaxy’s evolutionary history. However, in specific models like BHs3hl7v2L and BHs3hl6v10L, once the gas fraction diminishes to nearly zero, no new gas is available to fall into the galaxy, resulting in a permanent cessation of star formation.

Figures \,\ref{fig:age} through \ref{fig:metal} present the gradients of galaxy stellar age and metallicity for the redshift range $\mathrm{2 \lesssim z \lesssim 4}$.

\subsection{Black hole properties}
\label{sec:bh}
\begin{figure*}
 \includegraphics[width=0.9\textwidth]{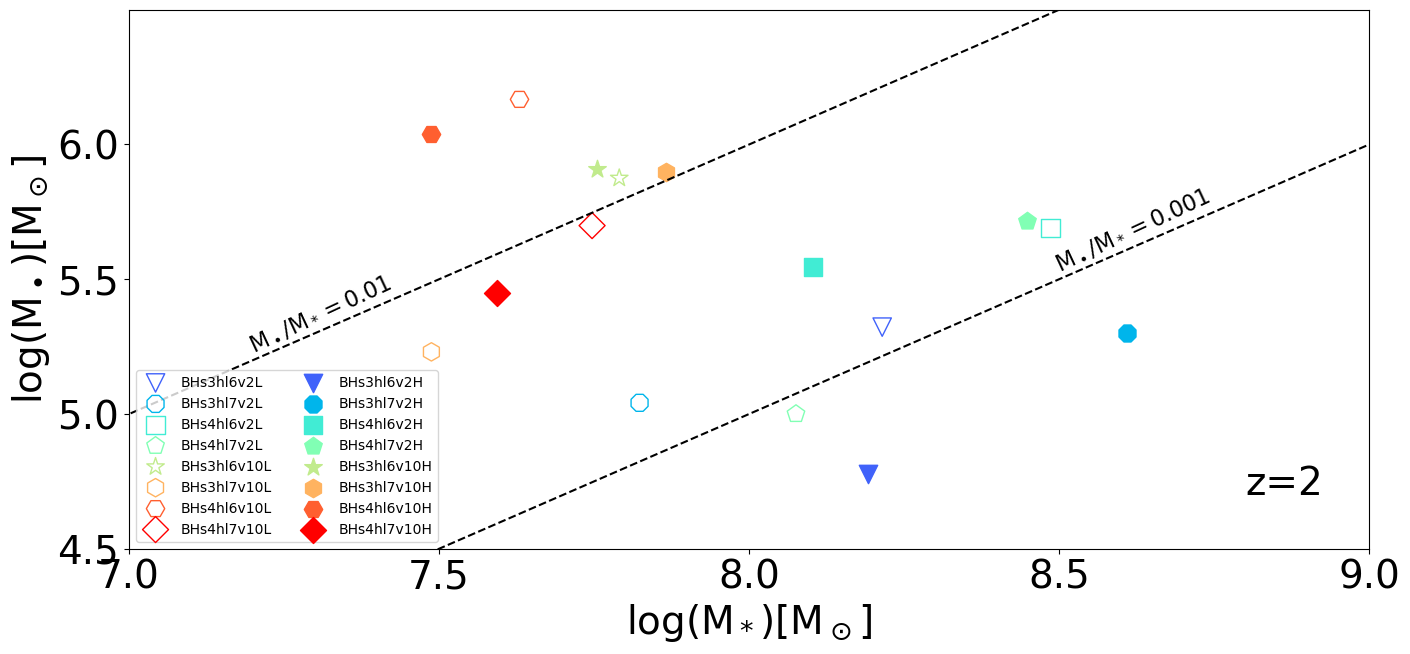}
    \caption{Relation between the black hole mass($\mathrm{M_\bullet}$) and its host galaxy stellar mass ($\mathrm{M_\star}$) at z=2. Models in low/high spin halos are labeled with hollow/solid markers respectively. Models with the same BH-related parameters in Table\,\ref{tab:BHsim} are labeled with the same marker shapes and colors.}
    \label{fig:star_vs_bh}
    \end{figure*}

This section examines the black holes (BHs) located within our final galaxy samples. The ratio of central black hole mass to host galaxy stellar mass, denoted as $\mathrm{M_\bullet-M_\star}$, is recognized as an important indicator of the formation history of supermassive black holes (SMBHs) \citep[]{vol03}. The ongoing debate centers around whether the $\mathrm{M_\bullet-M_\star}$ relationship evolves with redshift. In Figure\,\ref{fig:star_vs_bh}, we present the $\mathrm{M_\bullet-M_\star}$ relation for our sample at $\mathrm{z=2}$. Models are distinguished based on the spin of halos, with hollow markers representing low spin and solid markers indicating high spin. Models with identical BH-related parameters utilize the same marker shape and color. 

Overall, our analysis reveals that all galaxies in our sample host BHs with masses ranging approximately from $\mathrm{\sim10^{4.7}}$ to $\mathrm{\sim10^{6.2} M_\odot}$, which falls within the intermediate mass black hole (IMBH) category. The $\mathrm{M_\bullet-M_\star}$ ratio displays a scatter between 0.0003 and 0.03, which overlaps with the range reported in observational high-redshift quasar samples that have available JWST spectroscopic data from \citep[]{har23, yue23, ding23} and \citep[]{mai23}. 

Upon closer examination, our findings indicate that the $\mathrm{M_\bullet-M_\star}$ ratio separates our samples into two distinct branches: (1) models exhibiting high AGN wind velocities (designated by model names ending in v10L/H) show a ratio of approximately $\mathrm{M_\bullet/M_\star} \sim 0.01$; (2) models characterized by low AGN wind velocities (designated by model names ending in v2L/H) exhibit a ratio of around $\mathrm{M_\bullet/M_\star} \sim 0.001$. We do not observe significant differences in BH mass growth histories based on halo spin. However, three BH parameters in our models—the initial seed BH mass ($\mathrm{M_{seed}}$), the minimum halo mass for BH seeding ($\mathrm{M_{HaloMin}}$), and the AGN wind velocity for BH feedback ($\mathrm{v_{wind}}$)—have a considerable impact on BH growth.

As anticipated, models with smaller values of $\mathrm{M_{HaloMin}}$ tend to produce BHs earlier. When compared to models that differ only in specific BH parameters, variations in both $\mathrm{M_{seed}}$ and $\mathrm{M_{HaloMin}}$ typically yield differences in BH masses of approximately $\mathrm{2-7}$ times as early as $\mathrm{z\sim 4}$. The AGN wind velocity ($\mathrm{v_{wind}}$) emerges as the most pivotal factor influencing the rate of BH growth. An increase in $\mathrm{v_{wind}}$ favorably affects the BH accretion rate, resulting in a steeper mass growth curve until the curve flattens again as the BH consumes the available gas reservoir.

\begin{figure*}
 \includegraphics[width=0.9\textwidth]{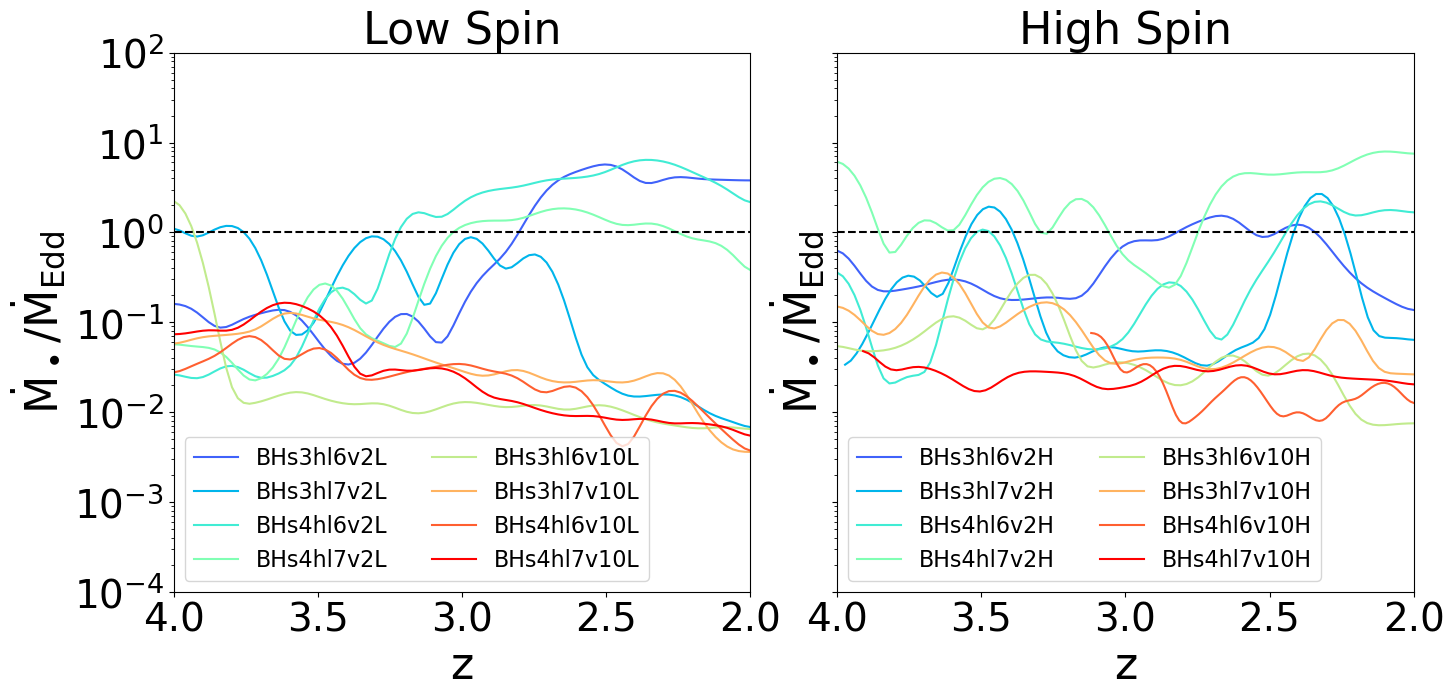}
    \caption{Evolution of the BH accretion rate Eddington ratio for galaxies in low spin (left) and high spin (right) host halos, from redshift $\mathrm{2 \ltorder z \ltorder 4}$. All models in Table\,\ref{tab:BHsim} are labeled with different colors.}
    \label{fig:bhedd}
    \end{figure*}

Figure\,\ref{fig:bhedd} illustrates the evolution of the black hole (BH) accretion Eddington ratio, defined as $\mathrm{ \mathrm{\dot{M}_\bullet/\dot{M}_{Edd}}}$, where $\mathrm{\dot{M}_{Edd}\approx2.38\frac{M_\odot}{yr}(\frac{M_\bullet}{10^8M_\odot})}$. At higher redshifts, specifically within the range $\mathrm{3 \ltorder z \ltorder 4}$, most of our models maintain a relatively constant Eddington ratio, approximately within the limits $\mathrm{0.01 \ltorder \dot{M}_\bullet/\dot{M}_{Edd} \ltorder 0.1}$. However, at redshifts lower than $\mathrm{z\sim 3}$, the behavior of models in low spin halos diverges from that of models in high spin halos: 

(1) In low spin halos, galaxies exhibiting low AGN (active galactic nucleus) velocities tend to experience an accretion rate boost, which can result in super-Eddington ratios ($\mathrm{\dot{M}_\bullet/\dot{M}_{Edd}\gtorder 1}$). Conversely, galaxies with high AGN velocities exhibit a decreasing trend, ultimately approaching a lower accretion ratio of approximately $\mathrm{\dot{M}_\bullet/\dot{M}_{Edd}\sim 5\times10^{-3}}$.

(2) In high spin halos, the situation is comparatively straightforward. The average Eddington ratio across each model remains relatively stable as redshift decreases. Nevertheless, variability in the accretion Eddington ratio persists. A notable observation is that models with low AGN wind velocities exhibit significant fluctuations in their accretion rates, with variations reaching factors of up to 100, occasionally surpassing the Eddington accretion rate at peak values. In contrast, models with high AGN wind velocities do not exhibit such pronounced fluctuations and consistently remain below the Eddington accretion limit.

\subsection{Gini/$\mathrm{M_{20}}$ coefficient}
\label{sec:gini}
The Gini/$\mathrm{M_{20}}$ coefficient is a widely accepted metric for describing the relative concentration of light within a galaxy, exhibiting insensitivity to the distribution of light within its central region. This coefficient is frequently employed as an indicator of mergers, a usage substantiated by its sensitivity to most face-on systems within the local universe \citep{tho15}. The definitions of Gini and M20, as proposed by \citep{abr03} and \citep{lotz04}, are given as follows:

\begin{equation}
Gini = \frac{1}{\bar{|X|} n (n-1)} \sum^n_i (2i - n -1) |X_i|
\end{equation}
where $\mathrm{\bar{|X|}}$ denotes the average flux value, $\mathrm{n}$ represents the total number of pixels in the image, and $\mathrm{|X_i|}$ is the flux value for each pixel, ordered by brightness in the summation.

\begin{eqnarray}
M_{20} \equiv {\rm log10}\left(\frac{\sum_i M_i}{M_{tot}}\right) & {\rm while } & \sum_i f_i <  0.2 f_{tot}
\end{eqnarray}
where $\mathrm{M_{tot}}$ represents the total second-order moment and $\mathrm{f_{tot}}$ denotes the total flux of the galaxy pixels identified by the segmentation map. Here, $\mathrm{M_{i}}$ and $\mathrm{f_i}$ refer to the second-order moment and flux for each pixel $\mathrm{i}$, respectively, such that $\mathrm{f_1}$ is the brightest pixel, $\mathrm{f_2}$ is the second brightest pixel, and so on.

\begin{figure}
 \includegraphics[width=0.48\textwidth]{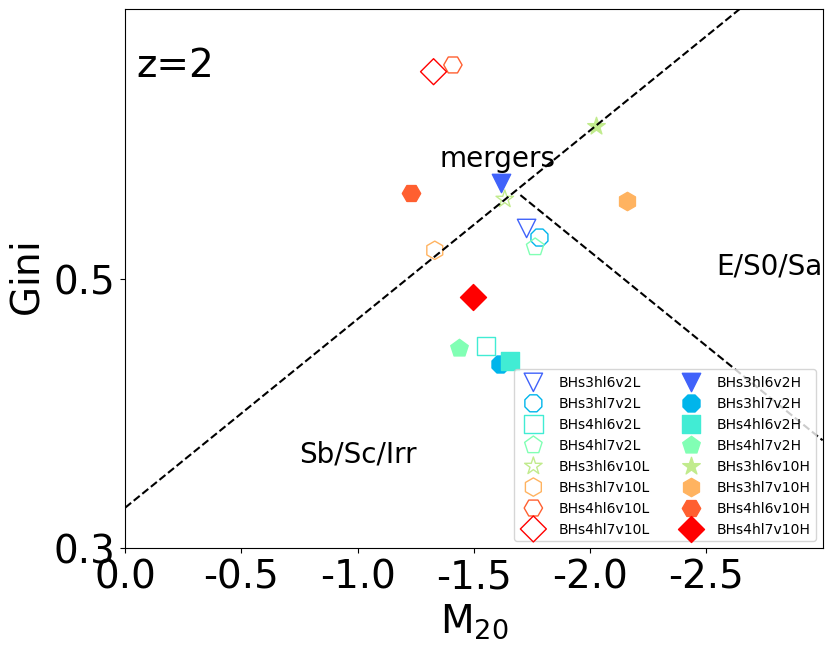}
    \caption{ Distribution of Gini and M20 parameters for all our models at final redshift $\mathrm{z=2}$. Models in low/high spin halos are labeled with hollow/solid markers separately. Models with the same BH-related parameters in Table\,\ref{tab:BHsim} are labeled with the same marker shapes and colors. Dashed lines separate the regimes into mergers/ellipticals/spirals morphological types (\citealp[]{lotz08}). }
    \label{fig:gini_m20}
    \end{figure}

In Figure\,\ref{fig:gini_m20}, we present the Gini coefficient compared to $\mathrm{M_{20}}$ at $\mathrm{z=2}$ for all the galaxies in our sample. All measurements have been conducted after orienting the galaxies face-on. Models associated with low and high spin halos are indicated with hollow and solid markers, respectively. Models sharing the same black hole (BH)-related parameters are represented with identical marker shapes and colors. The dashed lines delineate different morphological regimes; mergers are located above the dashed line, ellipticals are found in the right region, and spirals lie below the dashed line. We have adopted the following classification criteria based on the HST Survey of the Extended Groth Strip (EGS) from \(0.2 \ltorder \mathrm{z} \ltorder 1.2\) (\citealp[]{lotz08}):  
\begin{eqnarray}
\begin{array}{lll}
{\rm Mergers:} &  G > -0.14\ M_{20} + 0.33  &   \\
\\
{\rm E/S0/Sa:} &  G \le  -0.14\ M_{20} + 0.33     &\\
 &   \&\  G > 0.14\ M_{20} + 0.80  & \\
 \\
{\rm Sb-Ir:} &  G  \le  -0.14\ M_{20} + 0.33  &\\
&   \&\ G \le 0.14\ M_{20} + 0.80&  \\
\end{array}
\end{eqnarray}

At \(\mathrm{z=2}\), half of our models are classified within the spiral regime, while the remainder predominantly fall into the mergers regime. Only the model BHs3hl7v10H is situated in the elliptical regime. No systematic bias is observed in the phase space as a function of host halo spins. Notably, AGN wind velocity appears to influence the Gini values; models characterized by high wind velocities tend to exhibit higher Gini coefficients. The distribution of \(\mathrm{M_{20}}\) appears to be more random within the range \(\mathrm{-2.3 \ltorder M_{20}\ltorder -1.0}\), showing no significant correlation with any BH-related parameters, such as initial mass or seeding time.

\begin{figure}
 \includegraphics[width=0.48\textwidth]{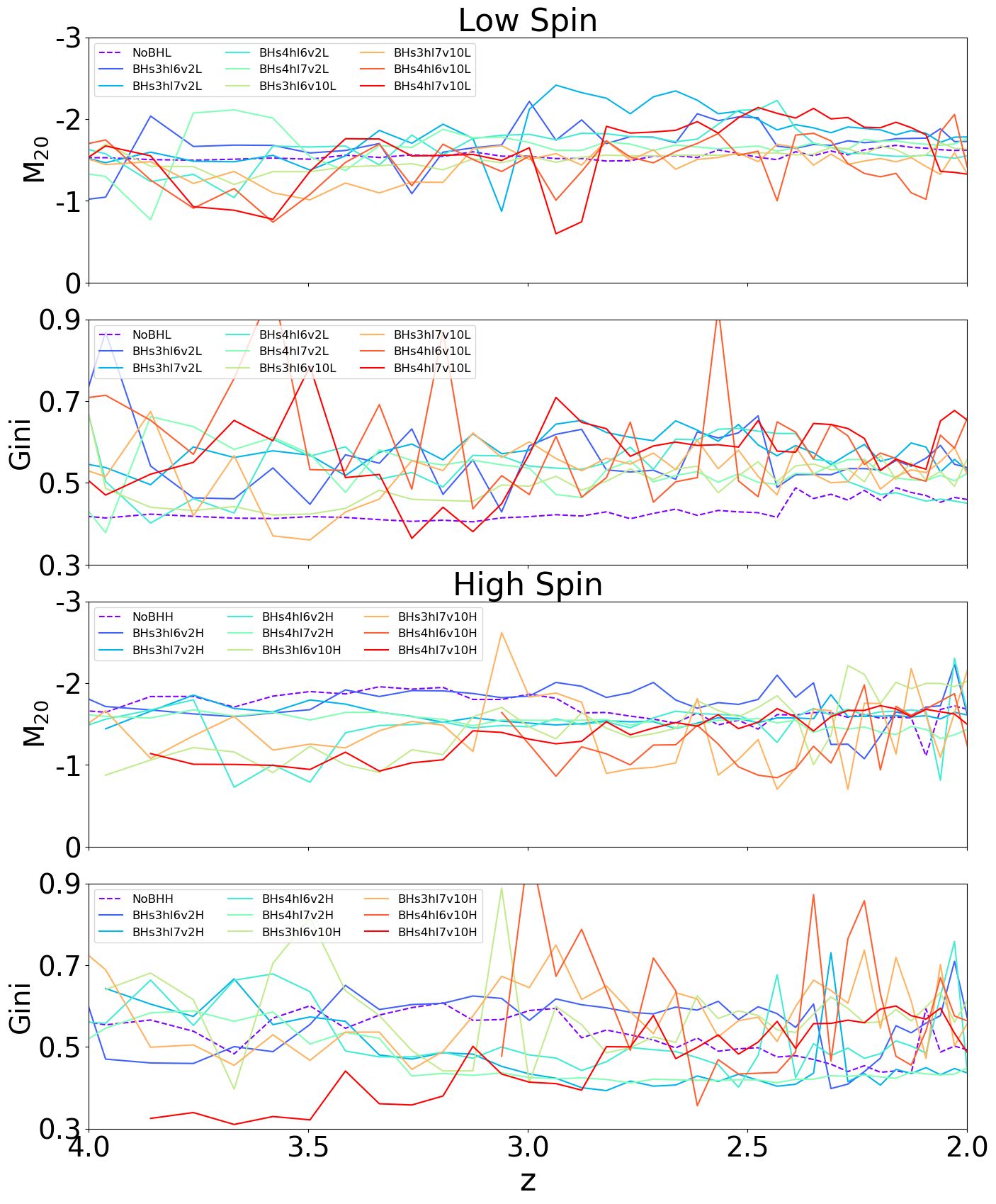}
    \caption{ Evolution of  Gini/M20 parameters for galaxies in low spin (top panels) and high spin (lower panels) host halos, from redshift z = 4 to 2. All models in Table\,\ref{tab:BHsim} are labeled with different colors.}
    \label{fig:gini_m20_z}
    \end{figure}

We also illustrate the Gini and \(\mathrm{M_{20}}\) coefficients for all galaxies as a function of redshift in Figure\,\ref{fig:gini_m20_z}. The models are categorized according to host halo spins (low spin at the top and high spin at the bottom). The \(\mathrm{M_{20}}\) values for each model do not demonstrate a clear trend of either increasing or decreasing; instead, they remain relatively constant, displaying some fluctuations within the range \(\mathrm{-2 \ltorder M_{20}\ltorder -1}\). A similar pattern is observed in the Gini coefficient, albeit with certain exceptions. For instance, the Gini values for model BHs4hl7v10H show a monotonically increasing trend from 0.3 to 0.5.

After applying the same morphological classification criteria utilized for galaxies at a final redshift of $z=2$, we identify all mergers based on the Gini-$\mathrm{M_{20}}$ indicator. To assess its effectiveness, we evaluate each galaxy classified as merging by examining actual nearby galaxies within a radius of 10 $\mathrm{kpc\,h^{-1}}$ from the primary galaxy in the snapshots. If the Gini-$\mathrm{M_{20}}$ analysis indicates that a galaxy is in a merger but no corresponding merging galaxy is identified in the snapshots, we categorize this instance as an artificial merger identification based on the Gini-$\mathrm{M_{20}}$ method. In Table\,\ref{tab:verify_merger}, we summarize the ratio of artificial mergers for all our galaxies spanning redshifts from $\mathrm{z=4}$ to $\mathrm{z=2}$, as determined by the Gini-$\mathrm{M_{20}}$ merger indicator. The results indicate that the Gini-$\mathrm{M_{20}}$ method is not an efficient tool for identifying mergers within our sample. Certain models, such as BHs4hl6v2L and BHs3hl6v2H, exhibit misidentification rates exceeding 90\%. Furthermore, we observed that in our galaxy sample, it is common for the Gini value to exceed the upper limit established for Ellipticals and Spirals at a fixed $\mathrm{M_{20}}$. Compared to typical low redshift galaxies, our high redshift galaxies ($z > 2$) tend to be more concentrated and many exhibit flattened bright centers. These characteristics likely contribute to the elevated Gini and $\mathrm{M_{20}}$ values, thereby resulting in confusion in the interpretation of the Gini-$\mathrm{M_{20}}$ indicator and leading to erroneous identifications.

\begin{table}
\caption{Artificial merger ratio of all our galaxies through $\mathrm{z=4-2}$ depends on Gini-$\mathrm{M_{20}}$ merger indicator criteria (\citealp[]{lotz08}).} 
\label{tab:verify_merger}
\vspace{-.1cm}
\centering
\resizebox{1.\columnwidth}{!}{%
\begin{tabular}{cc||cc}
\hline
\hline
 Model No. & Artificial merger ratio & Model No. & Artificial merger ratio \\
\hline
    NoBHL     & 0.00  & NoBHH &   0.71\\
\hline
  
    BHs4hl6v2L    & 0.91 & BHs4hl6v2H    & 0.16 \\

\hline   
    BHs4hl7v2L    & 0.04 & BHs4hl7v2H    & 0.78\\
\hline 
    BHs4hl6v10L    &0.03 & BHs4hl6v10H    & 0.18 \\

\hline   
   BHs4hl7v10L    & 0.09 & BHs4hl7v10H    &0.67  \\
\hline 
  
    BHs3hl6v2L     & 0.40 & BHs3hl6v2H     & 0.90  \\

\hline   
    BHs3hl7v2L   & 0.22 &  BHs3hl7v2H   & 0.33  \\
\hline 
   BHs3hl6v10L     & 0.85 & BHs3hl6v10H     & 0.01  \\

\hline   
    BHs3hl7v10L     & 0.58 & BHs3hl7v10H     & 0.00  \\
\hline   
\hline
\end{tabular}
}
\end{table}

\subsection{CAS statistics}
\label{sec:cas}
\begin{figure}
 \includegraphics[width=0.48\textwidth]{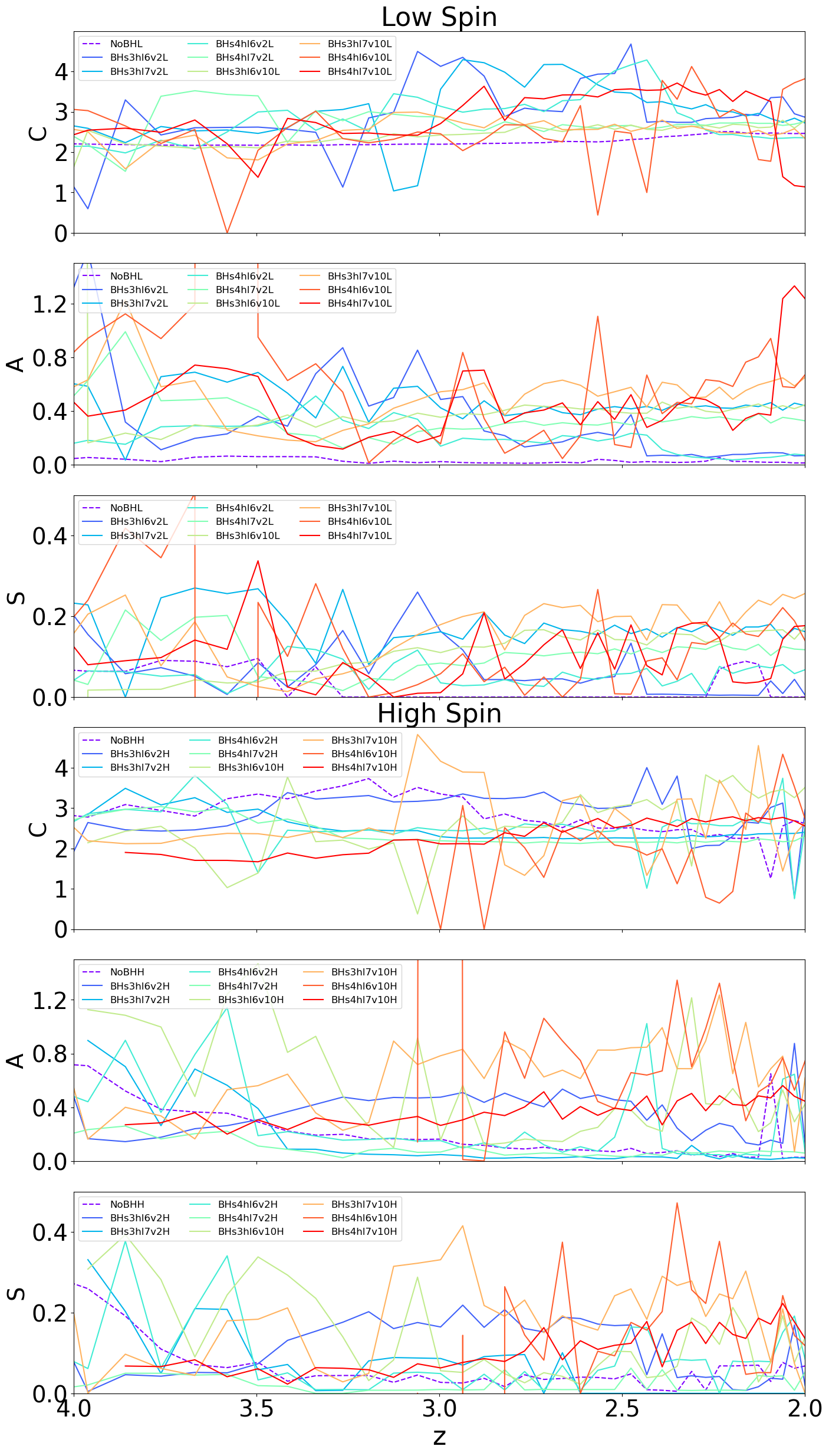}
    \caption{ Evolution of  CAS parameters for galaxies in low spin (top panels) and high spin (lower panels) host halos, from redshift z = 4 to 2. All models in Table\,\ref{tab:BHsim} are labeled with different colors.}
    \label{fig:cas_z}
    \end{figure}

The CAS system is a widely adopted non-parametric technique for galaxy morphological classification. In this system, "C," "A," and "S" represent concentration, asymmetry, and smoothness, respectively. The definitions of these parameters are provided below:\\
\\
According to \citealp[]{ber00}, concentration is defined as the ratio of the circular radii that encompass 20\% and 80\% of the total flux:
\begin{equation}
C = 5\ \rm{ log10}\left(\frac{r_{80}}{r_{20}}\right)
\end{equation}
where $\mathrm{r_{80}}$ and $\mathrm{r_{20}}$ are the circular apertures that contain 80\% and 20\% of the total flux, respectively. \\
\\
The asymmetry parameter $\mathrm{A}$ quantifies the extent to which the light distribution of a galaxy is rotationally symmetric. This parameter is determined by subtracting the galaxy image rotated by 180 degrees from the original image, as indicated by \citealp[]{abr95,wu99,con00}:
\begin{equation}
A = \sum_{i,j} \frac{ | I(i,j) - I_{180}(i,j)|}{|I(i,j)|}
\end{equation}
Here, $\mathrm{I}$ represents the image of the galaxy, while $\mathrm{I_{180}}$ is the image that has been rotated by 180 degrees around the central pixel of the galaxy. The value of $\mathrm{A}$ is computed by summing over all pixels within 1.5 times the Petrosian radius from the center of the galaxy, which is defined by minimizing $\mathrm{A}$.\\
\\
The smoothness parameter $\mathrm{S}$ was developed by \citet{con03}, drawing inspiration from the work of \citet{tak99}, to quantify the degree of small-scale structure. First, the galaxy image is smoothed using a boxcar of a specified width, which is then subtracted from the original image. The resulting residual indicates the smoothness of the image by accounting for features such as compact star clusters. The smoothing scale length is selected as a fraction of the Petrosian radius:
\begin{equation}
S =  \sum_{i,j}\frac{ | I(i,j) - I_S(i,j)| } {|I(i,j)|}
\end{equation}
In this case, $\mathrm{I_S}$ refers to the image of the galaxy that has been smoothed with a boxcar width of 0.25 Petrosian radii. The summation for $S$ is performed over pixels within 1.5 Petrosian radii of the galaxy's center, while excluding pixels within a circular aperture equal to the smoothing length of 0.25 Petrosian radii.\\
\\
In Figure\,\ref{fig:cas_z}, we present the CAS parameters for all galaxies as a function of redshift. The models are categorized based on the spins of their host halos, with low spin models displayed at the top and high spin models at the bottom. For concentration, all galaxies are located within a narrow range of $\mathrm{1 < C < 4}$. In contrast, the asymmetry and smoothness parameters exhibit a more erratic distribution, lacking a monotonic correlation with time. Notably, all CAS parameters indicate that models without black holes consistently display the most stable curves, characterized by minimal fluctuations.

\subsection{Sersic fitting}
\label{sec:sersic}
The two-dimensional Sersic profile fittings, as described by \citealp{ser63}, are performed using the \textsc{STATMORPH} code on face-on stellar density projections. The Sersic profile is defined by the equation:
\begin{equation}
I(x,y)=I(R)=I_e\ exp[-b_n(\frac{R}{R_{half}})^\frac{1}{n}-1]
\end{equation}
Here, the constant $\mathrm{b_n}$ is determined such that $\mathrm{R_{half}}$ encompasses half of the total flux, and it can be computed numerically. Both the Sersic index $\mathrm{n}$ and the half-mass radius $\mathrm{R_{half}}$ are treated as free parameters in this fitting process. A Sersic index of $\mathrm{n=1}$ signifies an exponential disk, characteristic of spiral galaxies, while $\mathrm{n=4}$ corresponds to a de Vaucouleurs profile, typically indicative of elliptical galaxies. Generally, a higher value of $\mathrm{n}$ suggests a greater concentration towards the center.

\begin{figure}
 \includegraphics[width=0.49\textwidth]{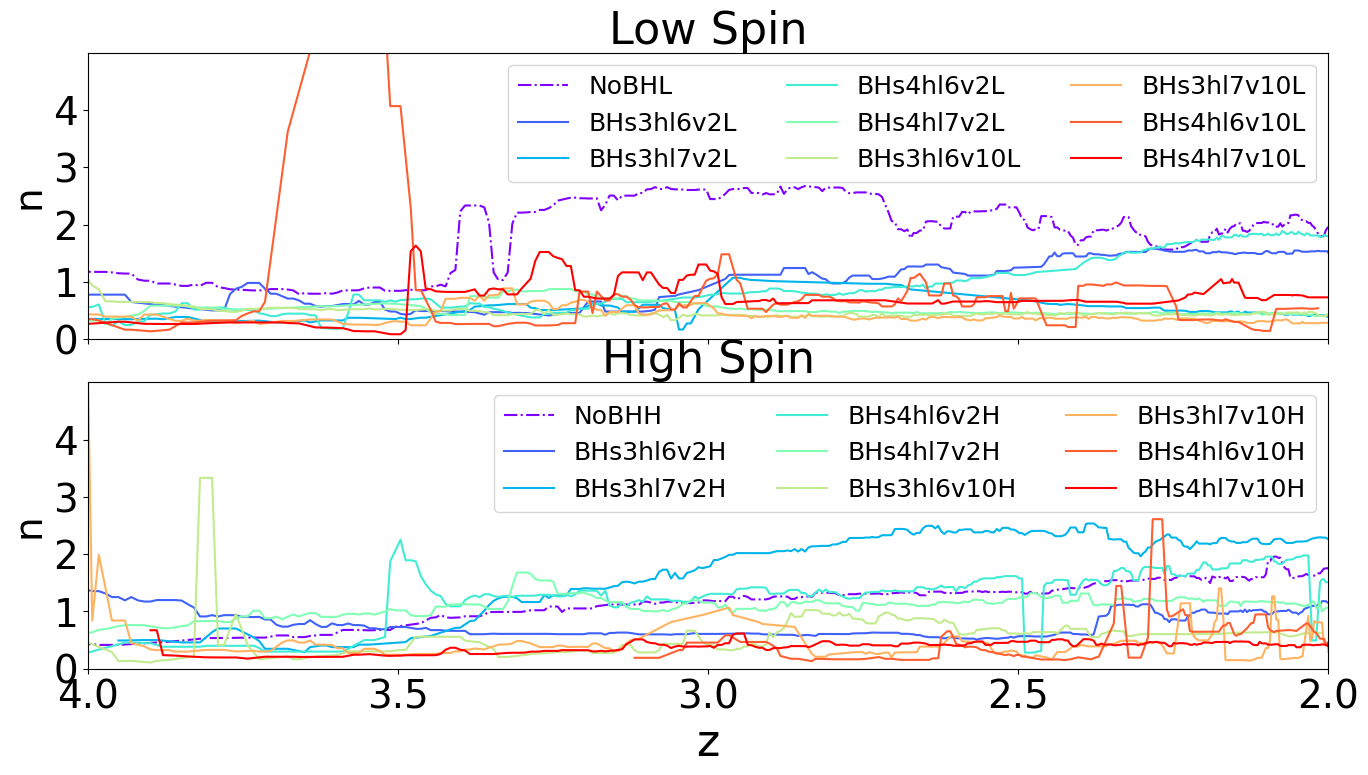}
    \caption{Evolution of Sersic index for the face-on surface density images of galaxies in low spin (top panel) and high spin (lower panel) halos, from redshift z = 4 to 2. All models in Table\,\ref{tab:BHsim} are labeled with different colors.}
    \label{fig:sersicn}
    \end{figure}

The evolution of the Sersic index $\mathrm{n}$ is depicted in Figure \ref{fig:sersicn}. Excluding fluctuations with substantial amplitudes, the values of $\mathrm{n}$ range from 0.3 to 3.0. Several models, such as BHs3hI6V2L and NoBHH, exhibit a monotonically increasing trend over time. In contrast, many other models maintain a consistently low index, remaining below $\mathrm{n < 1}$, such as BHs3hI6v10L, BHs3hI7V10L, and BHs4hI7V10H. Notably, the model without a black hole in a low spin halo (NoBHL) starts with a constant Sersic index around $\mathrm{n \sim 1}$ until approximately $\mathrm{z \sim 3.4}$, after which it experiences a significant increase, stabilizing at a higher index of approximately $\mathrm{n \sim 2}$. 

Additionally, we attempted to fit a double Sersic model, with the expectation that one Sersic profile would represent a bulge component and another would represent a disk component. However, all our models demonstrated no notable improvement in fitting quality, indicating that a single Sersic component fitting is sufficient.

Overall, the results for the Sersic index suggest that all of our studied galaxies exhibit characteristics of both exponential disks and elliptical galaxies. Although no prominent bulge components were detected, some galaxies, particularly those with black holes, display flattened profiles in their central regions.

\subsection{Rotational support}
\label{sec:rot}

\begin{figure}
 \includegraphics[width=0.49\textwidth]{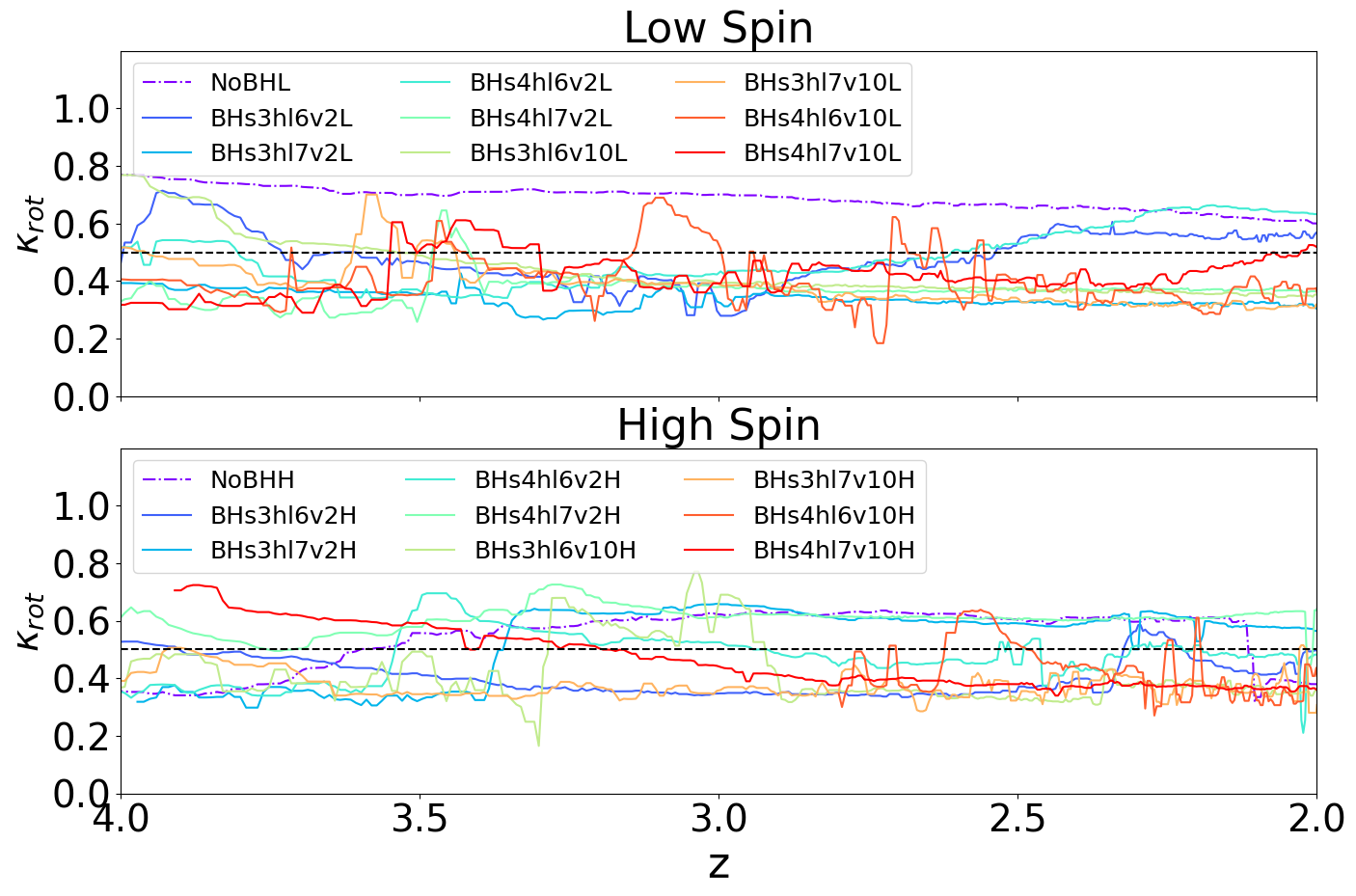}
    \caption{Evolution of  rotational support  for the galaxies in low spin (top panel) and high spin (lower panel) halos, from redshift z = 4 to 2. All models in Table\,\ref{tab:BHsim} are labeled with different colors.}
    \label{fig:krot}
    \end{figure}

The fraction of kinetic energy invested in ordered rotation, as defined by \cite{sales12}, is given by
  \begin{equation}
    \kappa_{\rm rot} = \frac{K_{\rm rot}}{K} = \frac{1}{K}\sum\frac{1}{2}\left(\frac{J_z}{R}\right)^{2}.
  \end{equation}
In this context, the z-direction aligns with the angular momentum of the stellar content of the galaxy. Figure \ref{fig:krot} illustrates the evolution of $\mathrm{\kappa_{\rm rot}}$. In both models without black holes, galaxies exhibit rotational support ($\mathrm{\kappa_{\rm rot} > 0.5}$) for the majority of the time. With the exception of a few peak fluctuations, models experiencing significant AGN wind velocities typically show $\mathrm{\kappa_{\rm rot} \leq 0.5}$. Additionally, in galaxies characterized by consistently low gas fractions (such as BHs3hl6v10L and BHs4hl7v10H, as shown in Figure \ref{fig:gas_frac}), the $\mathrm{\kappa_{\rm rot}}$ values tend to monotonically decrease from levels above 0.5 to approximately 0.35, before stabilizing until $\mathrm{z = 2}$.

\subsection{Mergers and tidal interactions}
\label{sec:merger}
Prior simulations of major mergers, alongside observations of local gas-rich merger events \citep[e.g.,][]{mh96,cox06,sch82}, have demonstrated that the merger process can initiate violent starbursts and facilitate the transformation of disk galaxies into spheroids. The star formation history of a galaxy is closely linked to its morphological evolution; thus, we meticulously examine all snapshots to verify galaxy interaction events.

\begin{figure}
 \includegraphics[width=0.48\textwidth]{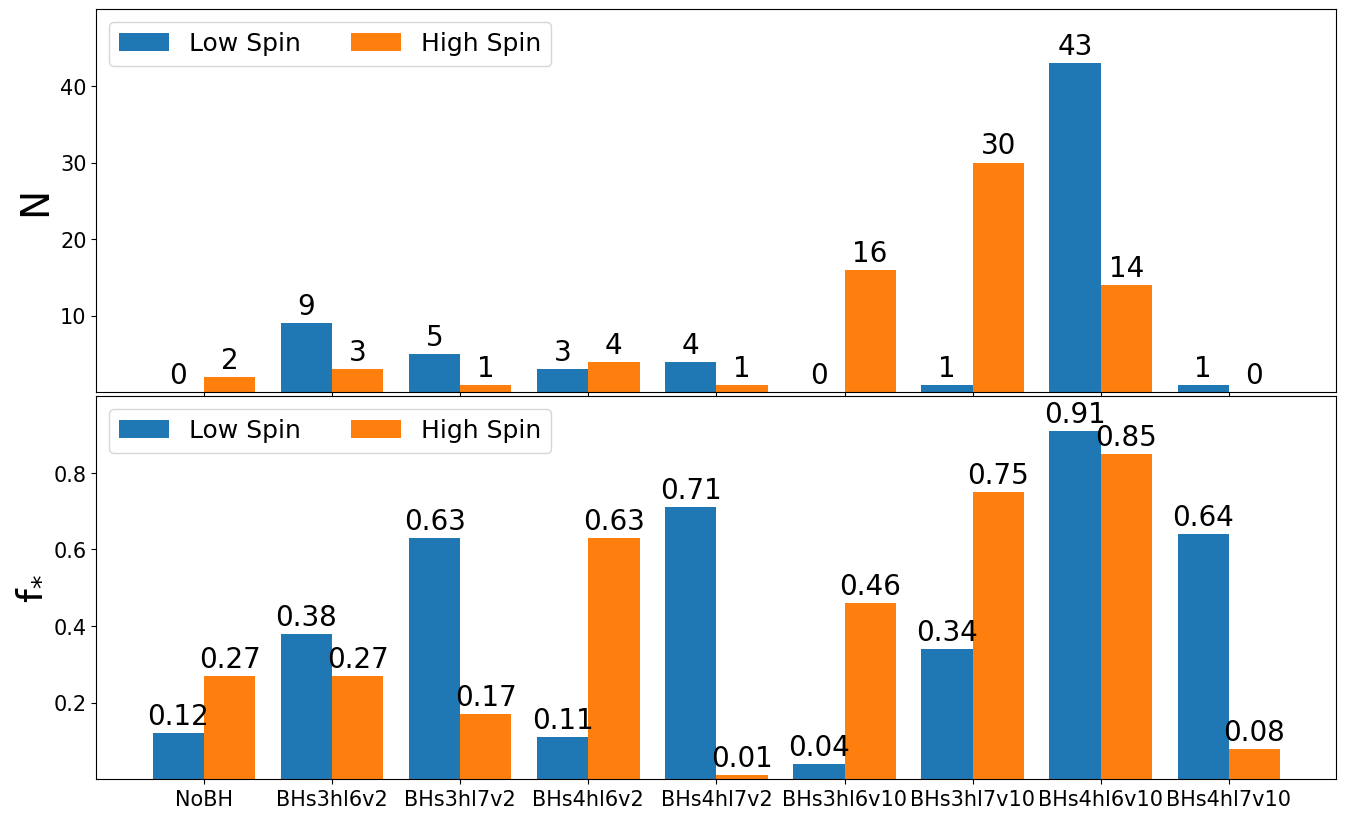}
    \caption{Merger statistics: Top panel --- major merger counts for all models from $\mathrm{2 \ltorder z \ltorder 4}$; bottom panel --- stellar mass contribution from mergers since galaxies reach 10\% of their final mass.}
    \label{fig:merger}
    \end{figure}

We define a major merger according to two criteria: (1) the stellar mass ratio of the colliding galaxies must be less than 3, and (2) at least 50\% of the stellar mass of the merging galaxy must ultimately merge into the main galaxy. In cases involving multiple merger events, each merger is assessed separately based on the aforementioned criteria. Galaxy interaction events that meet the first criterion but not the second are classified as close flybys. Furthermore, by tracing all stellar particles from \( \mathrm{2 < z < 4 }\), we calculate the percentage of stellar mass contributed by merger events (both major and minor) in the final galaxies. The summarized statistics are illustrated in Figure\,\ref{fig:merger}. Models exhibiting high AGN wind velocities show a significantly increased likelihood of undergoing major mergers, particularly for galaxies situated within high spin halos. This observation is congruent with the finding that these galaxies also exhibit higher percentages of stellar mass, highlighting the crucial influence of mergers on the mass growth of these galaxies.

\begin{figure}
 \includegraphics[width=0.48\textwidth]{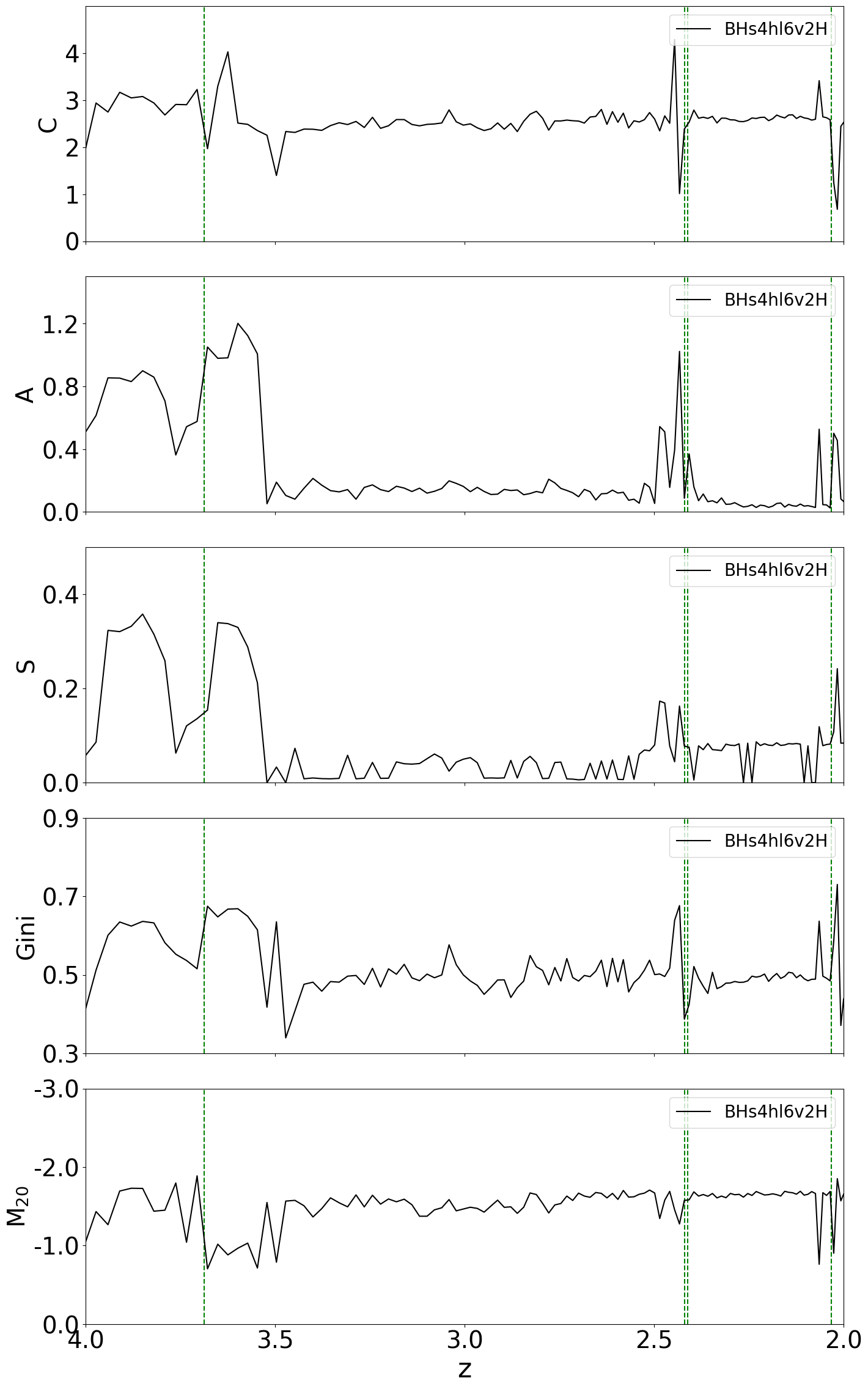}
    \caption{Timeline of CAS and Gini/$\mathrm{M_{20}}$ parameters evolution for representative model BHs4hl6v2H, from redshift z = 4 to 2. Green vertical dash lines indicate major mergers.}
    \label{fig:merger_example}
    \end{figure}

Additionally, we evaluate the impact of major mergers on various galaxy morphological parameters (e.g., model BHs4hl6v2H, as depicted in Figure\,\ref{fig:merger_example}). The vertical dashed lines delineate the termination times for major mergers. Clearly, major mergers consistently induce rapid, violent changes on short timescales. For instance, noticeable spikes correspond to mergers occurring at \( \mathrm{z \sim 2.4 }\) and \( \mathrm{z \sim 2.1 }\). However, the parameter values typically recover to their prior levels at lower redshifts (\(\mathrm{ z < 3 }\)). Consequently, for these galaxies at lower redshifts, major mergers do not exhibit a substantial impact on the overarching evolutionary trend across nearly all morphological parameters.

\subsection{Surface photometry and morphology}
\label{sec:photo}
\begin{figure}
 \includegraphics[width=.48\textwidth]{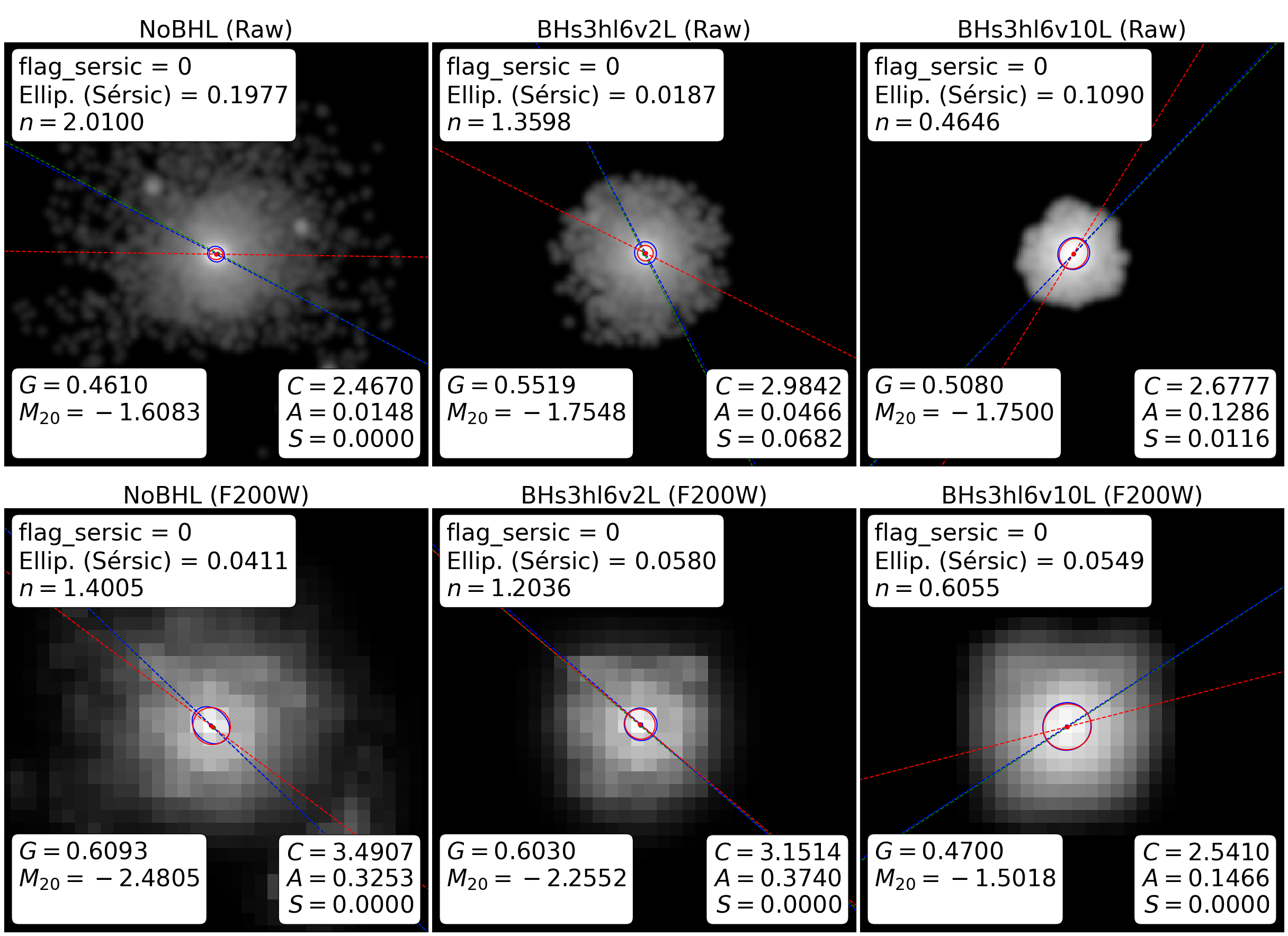}
    \caption{Morphology measurement for representative galaxy group includes model NoBHL, BHs3hl6v2L and BHs3hl6v10L. The first model is a galaxy without a BH and the latter two are the same galaxy with the same BH seeding initial mass and time but different AGN wind velocities. The face-on images from the raw stellar density (top)
and synthetic luminosity (bottom) are illustrated separately. The surface luminosity images have been obtained by post-processing the stellar light with a 3-D radiation transfer, then redshifted, pixelized, and convolved with the JWST PSF. The dust absorption has been implemented as well. The JWST filter F200W have been used. Morphological measurements and fittings are also calculated by \textsc{STATMORPH}, including the Gini/$\mathrm{M_{20}}$, CAS and single Sersic profile. The galaxy center, major axis and half-light radius of the asymmetry and fitting Sersic model are colored with blue and red separately.}
    \label{fig:photo}
    \end{figure}

We have conducted morphology calculations for all the photometric images of our face-on galaxies. Figure\,\ref{fig:photo} presents a representative group of galaxies at $\mathrm{z=2}$, including models NoBHL, BHs3hl6v2L, and BHs3hl6v10L. The first model, NoBHL, represents a galaxy without a black hole (BH), while the latter two models incorporate the same initial mass and seeding time for the BH, differing solely in their AGN wind velocities. The top row of the figure displays face-on images derived from the raw stellar density projections, whereas the bottom row showcases their synthetic luminosity counterparts. These surface luminosity images are generated using a mosaic of JWST F200W filters, as described in Section\,\ref{sec:light}. The images were obtained through a post-processing technique that involved 3-D radiation transfer (accounting for dust absorption), followed by redshifting, pixelization, and convolution with the point spread function (PSF). 

All morphology measurements, including Gini/$\mathrm{M_{20}}$, CAS, and Sersic fitting, are labeled in both panels. This figure is provided for comparative purposes, and we will only comment generally on its implications for galactic morphology. In the raw images (top), we observe that the model without a BH exhibits the largest scale and possesses more intricate structures. The model with low AGN wind velocity (middle) is more compact, although it remains larger than the model with high AGN wind velocity (right). While all three models display bright central structures, the gradients in radial intensity vary: stronger AGN feedback (higher AGN wind velocity) correlates with a flatter radial slope of inner intensity (as evidenced by the Sersic index).

In contrast to the raw density images, the surface luminosity images reveal pronounced smearing effects due to the limitations imposed by instrument pixel size and PSF convolution. Consequently, this effect results in a loss of small-scale structural features, potentially leading to a substantial increase in the half-light radius by flattening the light distribution. Furthermore, this can affect the calculation of the Petrosian radius, which is crucial for determining the appropriate aperture for morphology measurements. As a result, more than half of the photometric images in our sample exhibit increased concentration but reduced smoothness. However, we did not observe a consistent trend in other morphology parameters. Finally, we also attempted double Sersic fittings across all photometric images. Similar to the density images, none of our galaxies displayed significant improvement in fitting quality upon the addition of an extra Sersic component. Almost all models were adequately represented using a single Sersic profile, with the exception of model BHs4hl6v2L.

\section{Discussion}
\label{sec:disscussion}
In this study, we utilize cosmological hydrodynamical simulations to investigate the morphology of dwarf galaxies harboring intermediate-mass black holes (IMBHs) at a final redshift of \(\mathrm{ z=2} \). The specifics of our simulation setup are elaborated in Section\,\ref{sec:modeling}. We hypothesize that the evolution of IMBHs within dwarf galaxies is influenced by several key factors: the presence of gas, the efficiency of feedback mechanisms, and the cosmic environment. These factors collectively determine the likelihood of mergers and the supply of fresh gas, which in turn dictates the co-evolution of the dwarf galaxy's morphology and the central black hole.

To examine this hypothesis, we designed a series of simulations featuring galaxies situated within similar mass dark matter (DM) halos, each constrained by varying growth dynamics and feedback parameters of the central black holes. Our primary focus is on analyzing the morphological characteristics of these galaxies. We employ non-parametric morphological diagnostics, such as the Gini-M20 \citep{lotz04} and CAS \citep{con03} statistics, calculated from 2D surface density images using the \textsc{STATMORPH} code. Additionally, we perform 2D Sersic fits. Post-radiative transfer calculations are conducted using the \textsc{SKIRT} code, allowing us to generate idealized synthetic James Webb Space Telescope (JWST) images with filter F200W. We also repeat morphological measurements on these photometric images.

Our findings indicate that, despite all galaxies evolving within similar mass DM halos (~\(\mathrm{10^{10} M_\odot}\)), their final stellar masses are significantly influenced by specific parameters of their black hole models rather than the spin of their host halos (\(\mathrm{\lambda}\); see Figure\,\ref{fig:stellar_mass}). Specifically, compared to models without black holes (BHs), those with low AGN wind velocities tend to generate galaxies of similar final masses (varying by a factor of 1-3). Conversely, the final stellar masses of models with high AGN wind velocities can be up to 1 dex lower than their no-BH counterparts. Other BH parameters, such as the initial mass and seeding time, appear to have negligible effects on the final stellar mass, though they do influence the history of galaxy growth. 

For example, the models designated BHs4hl6v10H and BHs4hl7v10H exhibit nearly identical black hole settings except for their seeding times. The former model seeds the BH earlier, resulting in a stellar mass of \(\mathrm{M_\bullet \sim 10^{7.5}M_\odot}\) by \(\mathrm{z=4}\), after which growth significantly slows. In contrast, the latter model experiences BH seeding at a later time and undergoes rapid mass growth between \(3< z < 4\) until stabilizing and halting growth similarly to the former model. A comparable trend is observed when comparing models with varying initial seeding masses for the BHs; lower initial masses require more time for growth, subsequently delaying the termination of rapid growth in their host galaxies.

Further analysis in conjunction with the evolution of gas fractions and star formation rates, as illustrated in Figures\,\ref{fig:gas_frac} and \ref{fig:sfr}, reveals a strongly correlated periodic rise and fall (see Figure\,\ref{fig:gasfrac_vs_sfr}). This suggests a scenario where the energy accumulated from central IMBH accretion regulates the cooling and accretion of cold gas in the host galaxies. The overall growth of the galaxies is determined by the relative strengths of inner AGN feedback and the available gas reservoir. When sufficient gas is present to fuel the central IMBH, it promotes the growth of \(\mathrm{M_\bullet}\) and enhances BH accretion, which in turn releases additional heating energy and generates outflows that hinder further gas accretion, consequently suppressing star formation in the host galaxies.

As the gas within the galaxy is swiftly depleted, both the central IMBH accretion and AGN feedback decrease, facilitating additional gas inflow until a new equilibrium is achieved. This dynamic may elucidate the monotonicity of the radial gradients of age and metallicity, as displayed in Figures\,\ref{fig:age} and \ref{fig:metal}, which could result from inside-out star formation quenching linked to negative AGN feedback. Recent observational studies provide compelling evidence for similar scenarios in supermassive black holes (SMBHs) (\citealp{wang24}). Our simulations suggest that a comparable phenomenon can also occur in dwarf galaxies with IMBHs, albeit with the caveat that some of our galaxies cease this periodic rebalancing process due to a lack of gas inflow and, consequently, stop growing at high redshifts. This stagnation may be attributed to the shallow gravitational potential of dwarf galaxies, which is insufficient to attract gas as effectively as more massive galaxies.

We have also investigated the growth and accretion rates of the intermediate-mass black holes (IMBHs) in our dwarf galaxies, as discussed in Section\,\ref{sec:bh}. Generally, all of our final galaxies exhibit comparable $\mathrm{M_\bullet-M_\star}$ ratios when compared to high-redshift observational samples. The final $\mathrm{M_\bullet-M_\star}$ ratio demonstrates a strong correlation with AGN wind velocity: models characterized by high velocities yield a ratio of $\mathrm{M_\bullet/M_\star \sim 0.01}$, while those with low velocities result in a ratio of $\mathrm{M_\bullet/M_\star \sim 0.001}$. The distinctions among other model parameters, such as host halo spin, initial seed masses of black holes (BHs), and timescales, are primarily discernible in the accretion evolution history depicted in Figure\,\ref{fig:bhedd}. In this figure, fluctuations in the history typically correlate with the instantaneous star formation rate and gas fraction levels. In light of the galaxy growth discussion presented earlier, we conclude that AGN feedback efficiency, which is governed by AGN wind velocity in our models, plays a significant role in determining the general properties of galaxies and BHs, such as mass, in an accumulative manner. Conversely, the initial mass and timing of BH seeding predominantly influence the timing of specific processes, which can be observed in the evolution history of galaxies and BHs.

In Sections\,\ref{sec:gini},\,\ref{sec:cas}, and \ref{sec:sersic}, we analyze the morphology of all models based on their face-on surface density images. We find that most of our galaxies fall within a phase range of $\mathrm{0.4 \ltorder Gini \ltorder 0.7}$ and $\mathrm{-2.5 \ltorder M_{20} \ltorder -1}$ for redshifts between $\mathrm{2 \ltorder z \ltorder 4}$. In comparison to galaxies from the EGS sample at redshifts of $\mathrm{0.2 \ltorder z \ltorder 1.2}$, many of our galaxies exhibit relatively higher Gini values at fixed $\mathrm{M_{20}}$. This discrepancy leads the empirical Gini-$\mathrm{M_{20}}$ indicator to erroneously classify these galaxies as undergoing mergers. The evolution of both Gini and $\mathrm{M_{20}}$ does not exhibit a simple general trend associated with any specific model parameters, including host halo spin, initial BH mass, time, and AGN wind velocity. However, we observe that models characterized by high AGN wind velocities tend to display significantly larger fluctuations in Gini amplitude. We attribute this phenomenon to the frequency of tidal events correlated with models exhibiting different intensities of AGN feedback, a topic we will address in further detail later. For CAS statistics, only the concentration index ("C") demonstrates a regular evolutionary trend akin to the Gini coefficient within a narrow range of $\mathrm{1 \ltorder C \ltorder 4}$. This trend can be explained by the well-documented correlation between Gini and concentration. We have attempted to fit all galaxy images using both single and double Sersic profiles. The results indicate that a single Sersic profile suffices for a good fit, and an additional bulge or disk component is unnecessary. Most galaxies with high AGN wind velocities maintain a Sersic index of approximately $n \sim 1$ since $z = 4$. In contrast, galaxies devoid of BHs or characterized by low AGN wind velocities may exhibit a monotonic increasing trend, typically culminating in a slightly higher Sersic index. Overall, the majority of our galaxies containing BHs present a well-fit single Sersic profile with an index of $\mathrm{n \ltorder 1}$, and no classical elliptical galaxies were identified.

The rotational support for all our galaxies is examined in Section\,\ref{sec:rot}. The typical fraction of rotational energy ranges from $\mathrm{0.3 \ltorder \kappa_{\rm rot} \ltorder 0.6}$. Consistent with the Sersic fitting results presented earlier, our galaxies are neither primarily supported by rotation, as typical disk galaxies are, nor by dispersion, as is the case for typical elliptical galaxies; instead, they represent a mixture of both support types. It is noteworthy that strong AGN wind feedback can significantly diminish rotational support. Galaxies devoid of BHs typically exhibit the highest levels of rotational support.

When discussing the evolution of galaxy morphology, mergers are an essential factor to consider. In Section \ref{sec:merger}, we examine mergers and tidal interactions occurring in our galaxy samples at redshifts ranging from \(\mathrm{2 \ltorder z \ltorder 4}\). The distinction between models characterized by low and high AGN wind velocities is clearly evident. The models with higher AGN wind velocities exhibit a greater frequency of major mergers and a larger percentage of mass contribution from mergers. This increased merger frequency in galaxies with elevated AGN wind velocities may contribute to the larger amplitude fluctuations observed in Gini evolution and the higher dispersion support levels previously reported. One possible explanation for the increase in merger frequency, as discussed in the literature (\citealp[]{sn09,silk13}), is that stronger AGN feedback, associated with high wind velocities in our models, compresses dense clouds and propagates gas clumps into the interstellar medium. This effect can locally enhance star formation and lead to the formation of smaller galaxies within the host halo at high redshift. In our simulations, we indeed observe more clumpy structures in galaxies with high AGN wind velocities, which can elevate the likelihood of mergers and other tidal interactions for central galaxies. Moreover, the differences in merger statistics between host halo spins are negligible. 

Lastly, by utilizing surface photometry with the JWST F200W filter and convolving with the point spread function (PSF), we find that nearly all of our galaxies exhibit larger half-light radii in luminosity images compared to density images. This phenomenon arises due to the comparable size of the compact, bright structures typically located in the central regions of our galaxies to the instrument's pixel size. Consequently, this results in slight modifications in our measurements of morphological parameters. However, we do not observe any consistent trends of increase or decrease in the measurements of surface density images, with the exception of concentration ("C") and smoothness ("S"). More than 50\% of our galaxies display higher concentration and lower smoothness in surface photometry images relative to surface density images, which can be readily explained by averaging effects resulting from resolution degradation due to pixelization and PSF convolution.

\section{Conclusion}
\label{sec:conclusion}
This study explores the role of intermediate-mass black holes (IMBHs) in shaping the morphology of dwarf galaxies at high redshift (z = 2) through cosmological hydrodynamical simulations. Our findings reveal that AGN feedback, particularly the strength of AGN-driven winds, plays a crucial role in regulating gas fractions, star formation rates, and the structural evolution of these galaxies. Models with higher AGN wind velocities lead to galaxies with lower stellar masses, flatter morphologies, and reduced rotational support. The morphology of IMBH-hosting galaxies is characterized by prominent central structures, low Sersic indices ($\mathrm{n < 2}$), and intermediate rotational support ($\mathrm{\kappa_{rot}}$ between 0.3 and 0.6). Traditional indicators like the Gini-M20 merger diagnostic prove ineffective for high-redshift systems, as their concentrated light distributions differ significantly from those at lower redshifts. Synthetic observations, modeled after James Webb Space Telescope (JWST) capabilities, suggest that pixelation and convolution effects can exaggerate observed galaxy sizes and concentrations. These results emphasize the intricate interplay between black hole evolution, feedback mechanisms, and the formation and morphology of dwarf galaxies, while also highlighting the challenges of interpreting high-redshift observations using low-redshift frameworks.

Considering the limited size of our galaxy samples, our results may be influenced by specific galaxy properties. Future observations from instruments such as JWST, Rubin, and the Roman Space Telescope (RST) will aid in addressing these limitations and enhance our understanding of the roles that IMBHs play in low-mass galaxies.

\begin{acknowledgements}
We thank Phil Hopkins for providing us with the latest version of the
code. We are grateful to Alessandro Lupi for his help with GIZMO and KROME,
and to Peter Behroozi for clarifications about ROCKSTAR. DB acknowledges insightful discussions with Isaac Shlosman via ZOOM. DB, DS \& AE acknowledges financial support from Millennium Nucleus NCN19 058 and support from the Centre for Astrophysics and
Associated Technologies CATA (FB210003). DRGS and AE thank for funding via ANID QUIMAL220002. DRGS thanks for funding via the Alexander von Humboldt - Foundation, Bonn, Germany. Simulations have
been performed using generous allocation of computing time on KULTRUN\footnote{http://www.astro.udec.cl/kultrun/} hybrid cluster in  Universidad de Concepci\'on and Leftraru cluster from the National Laboratory for High Performance Computing Chile (NLHPC)\footnote{https://www.nlhpc.cl/}.
\end{acknowledgements}


\begin{appendix}
\section{The gradient of galaxy stellar age and metallicity}

\begin{figure}

  \includegraphics[width=0.5\textwidth]{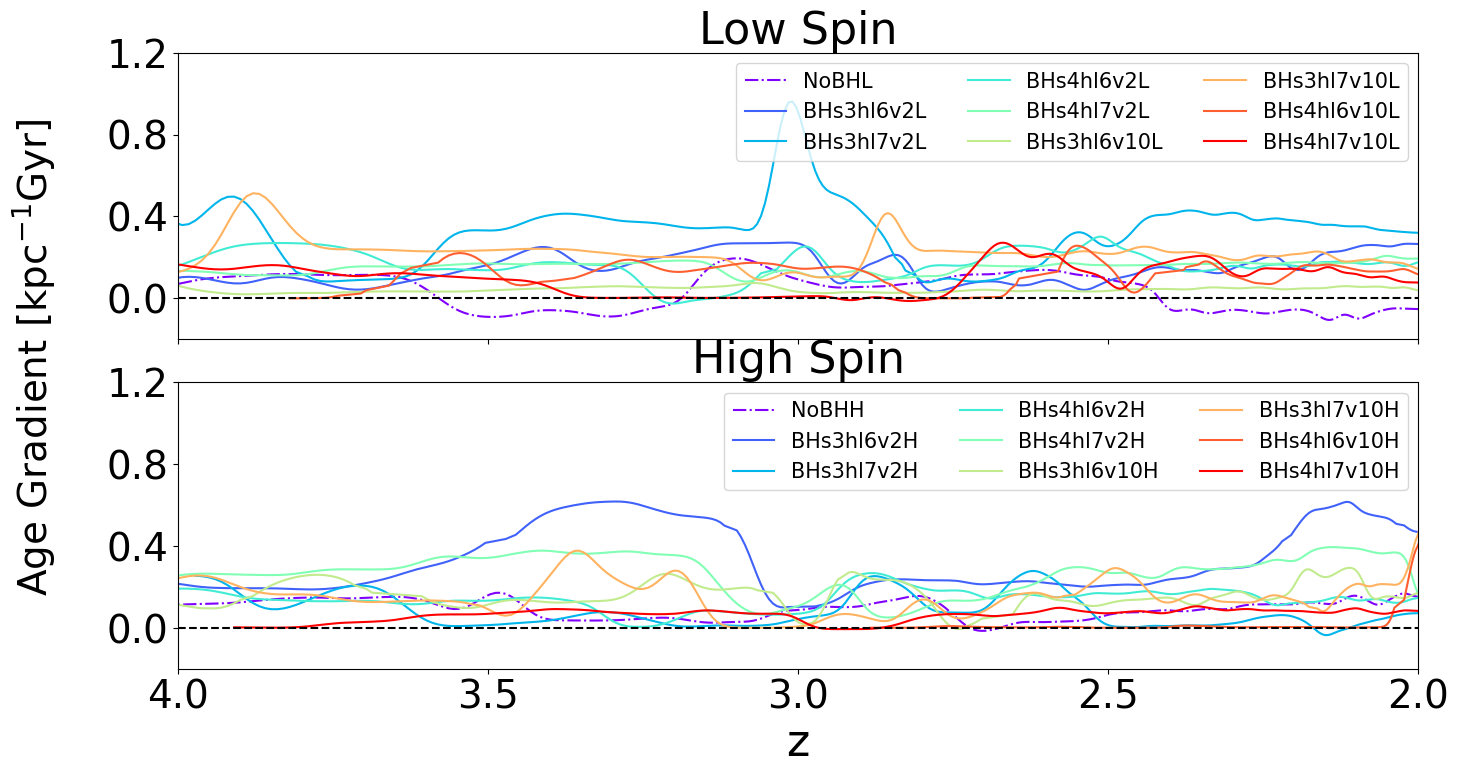}
    \caption{Evolution of stellar age gradient slope in low/high spin halo galaxy from redshift $z=4$ to 2. The black horizontal dashed line indicates an uniform distribution. All models in Table\,\ref{tab:BHsim} are labeled with different colors.}
    \label{fig:age}
    \end{figure} 

Radial age gradients serve as a cumulative record of the galaxy's stellar population build-up. Consequently, we examine the 1D stellar age radial profile and fit it with a straight line. We present the slope of the fitting line as a function of redshift in Figure \,\ref{fig:age}.  All models except NoBHL exhibit positive slopes. The slopes range from 0 to 0.5, with notable peaks at certain redshifts. We do not observe any discernible trend when comparing models with different BH-related parameters.

\begin{figure}

  \includegraphics[width=0.5\textwidth]{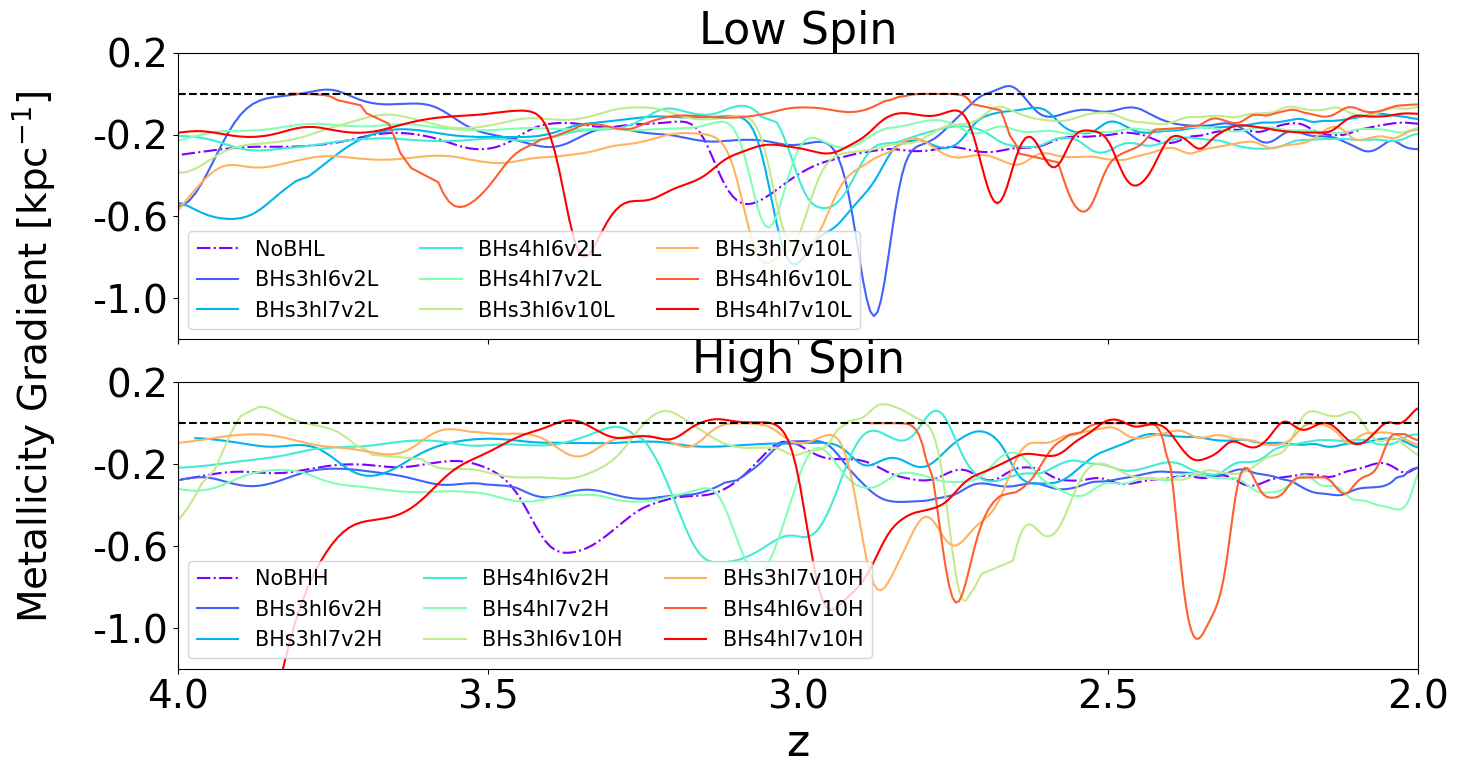}
    \caption{Evolution of stellar metallicity gradient slope in low/high spin halo galaxy from redshift $z=4$ to 2. The black horizontal dashed line indicates an uniform distribution. All models in Table\,\ref{tab:BHsim} are labeled with different colors.}
    \label{fig:metal}
    \end{figure} 

Similarly, the stellar metallicity gradient slope as a function of redshift is shown in Figure \,\ref{fig:metal}.  Contrary to the stellar age gradients, all models show negative slopes. Compared to the positive bumps in the age gradient slope profiles mentioned above, one can find the counterpart dips in the metallicity gradient slope profiles. However, there are no proportional relations for the fluctuation amplitudes.
\end{appendix}

\end{document}